\documentstyle[12pt,aaspp4,epsf]{article}

\def \etal {{\it et al. }}
\def \Mpc {h_{0.5}^{-1}{\rm Mpc}}

\def \eg {{\it e.g.,}}
\def \ie {{\it i.e.,}}

\def\deg{^\circ}

\def\ltsima{$\; \buildrel < \over \sim \;$} 
\def\lsim{\lower.5ex\hbox{\ltsima}}

\slugcomment{Ap.J., 2001, 561, 600}

\def\plottwo#1#2{\centering \leavevmode
\epsfxsize=.45\columnwidth \epsfbox{#1} \hfil
\epsfxsize=.45\columnwidth \epsfbox{#2}}

\def\plottwovert#1#2{\centering \leavevmode
\epsfxsize=.95\columnwidth \epsfbox{#1}\hfil
\epsfxsize=.95\columnwidth \epsfbox{#2}}

\def\plotthree#1#2#3{\centering \leavevmode
\epsfxsize=.3\columnwidth \epsfbox{#1} \hfil
\epsfxsize=.3\columnwidth \epsfbox{#2}  \hfil
\epsfxsize=.3\columnwidth \epsfbox{#3}}

\begin{document}
 
\title{Deprojection of Galaxy Cluster X-ray, Sunyaev-Zel'dovich Temperature
Decrement and Weak Lensing Mass Maps}
\author{S. Zaroubi\altaffilmark{1}, G. Squires\altaffilmark{2}, G. de
Gasperis\altaffilmark{3}, A. Evrard\altaffilmark{4}, Y.
Hoffman\altaffilmark{5}, and J. Silk\altaffilmark{3,6,7}
}

\altaffiltext{1}{Max Planck Institute for Astrophysics
Karl-Schwarzschild-Str 1, D-85740 Garching bei M\"unchen, Germany}

\altaffiltext{2}{California Institute of Technology, 1200 E. California
Blvd., M/C 105-24, Pasadena, CA 91125, U.S.A.}  

\altaffiltext{3}{Department of Astronomy, University of California,
Berkeley, CA 94720, U.S.A.}

\altaffiltext{4}{Physics Department, University of Michigan, Ann Arbor,
MI 48109, U.S.A} 

\altaffiltext{5}{Racah Institute of Physics, The Hebrew University, 
Jerusalem 91904, Israel}

\altaffiltext{6}{Center for Particle Astrophysics, University of
California, Berkeley, CA 94720, U.S.A.}  

\altaffiltext{7}{Oxford University Astrophysics, Nuclear Physics Lab.,
Dept. of Physics, Keble Rd, Oxford, OX1 3RH, UK} 

\begin{abstract}
 A general method of deprojecting two-dimensional images to reconstruct
the three dimensional structure of the projected object -- specifically
X-ray, Sunyaev-Zel'dovich (SZ) and gravitational lensing maps of rich
clusters of galaxies -- assuming axial symmetry (\cite{zshs98}), is
considered. Here we test the applicability of the method for realistic,
numerically simulated galaxy clusters, viewed from three orthogonal
projections at four redshift outputs. We demonstrate that the assumption
of axial symmetry is a good approximation for the 3D structure in this
ensemble of galaxy clusters. Applying the method, we demonstrate that a
unique determination of the cluster inclination angle is possible from
comparison between the SZ and X-ray images and, independently, between
SZ and surface density maps. Moreover, the results from these
comparisons are found to be consistent with each other and with the full
3D structure inclination angle determination. The radial dark matter and
gas density profiles as calculated from the actual and reconstructed 3D
distributions show a very good agreement. The method is also shown to
provide a direct determination of the baryon fraction in clusters,
independent of the cluster inclination angle.
\end{abstract}


\section{Introduction}
Recent years have witnessed significant improvements in the quality and
quantity of data from clusters of galaxies in all wavebands, from the
X-ray, through optical, to radio wavelengths. Together with advances in
the theoretical modeling of the physical processes in clusters, these
improvements call for new and more accurate methods of data analysis and
comparison with various theoretical predictions.

In particular, uncovering the 3D structure of clusters is of great
interest as it plays an important role in the determination of the
Hubble constant from X-ray and SZ measurements (see Appendix), cluster
mass and baryon fraction determination, and the underlying galaxy
orbit structure. In order to reconstruct the 3D geometry of clusters
researchers normally apply the spherical symmetry assumption, either within
a parametric (\eg\ Cavaliere \& Fusco-Femiano 1976) or non-parametric
(Fabian \etal 1981; and more recently Yushikawa \& Suto 1999)
framework. Although this simplifying assumption is very useful for
many problems, it can lead to significant biases in others, \eg\ the
Hubble constant determination (\eg\ Fabricant et al 1984).

Recently, a few studies have developed methods to take advantage of
the full two dimensional information retained in the cluster images
and extend the modeling of the underlying three dimensional structure
to have a more general, aspherical distribution (\eg\ Zaroubi \etal
1998; Reblinsky \& Bartelmann 1999). In particular, Zaroubi \etal
(1998; hereafter referred to as paper I) presented a model-independent
method of image deprojection for probing the 3D structure in clusters
of galaxies, assuming only that the underlying 3D structure has axial
symmetry.  The original testing of this method demonstrated that given
images of the cluster in the X-ray, radio and a map of the projected
total mass distribution, one could uniquely determine the inclination
angle of the system symmetry axis, and determine the underlying 3D
structure in the gas and dark matter (see also Grego \etal\
2000). They studied the stability of the inversion method in the
context of a specific analytical gas density profile: an axially
symmetric, elliptical isothermal model;
which is an extension of the widely used isothermal sphere model.

A few questions arise from the study in paper I: 1) Is the fundamental
assumption of axial symmetry valid for real galaxy clusters?  2) Does
substructure and dynamical evolution in galaxy clusters (as predicted to
be abundant by hierarchical structure formation models) permit a robust,
unique and stable numerical application of the method?  3) Can the
cluster symmetry axis inclination angle be uniquely determined for more
complex cluster gas/mass density distributions? 4) With realistic
expected resolution and signal-to-noise from current observations, can
the method be applied?

To probe these questions, we have performed a study of the inversion
algorithm on a set of numerically simulated galaxy clusters. The goal is
to test the axisymmetric deprojection algorithm, using as input a
representative realization of the cluster population.
We would like to emphasize at this stage that testing the
applicability of the technique to current data sets is beyond the
scope of the current paper. The method should be applied to each data 
set while taking into account its unique specifications, \eg\
resolution, scanning strategy, data acquisition domain (UV plane {\it
vs.} real space), etc. We intend indeed to tackle these issues for
specific data sets in a forthcoming paper.

The outline of this study is as follows.  In \S2 the deprojection method
is briefly reviewed, and we develop a new feature of the method that
allows an inclination angle-independent determination of the baryon
fraction in clusters.  In \S3, we briefly describe the input cluster
simulations.  In \S4, we show in detail the results for a single,
prototype simulated cluster. In \S5, we apply the method to the full
cluster simulation output. The paper is concluded with a discussion (\S6). 

\section{The Deprojection Method}

\begin{figure*}[ht]
\plotone{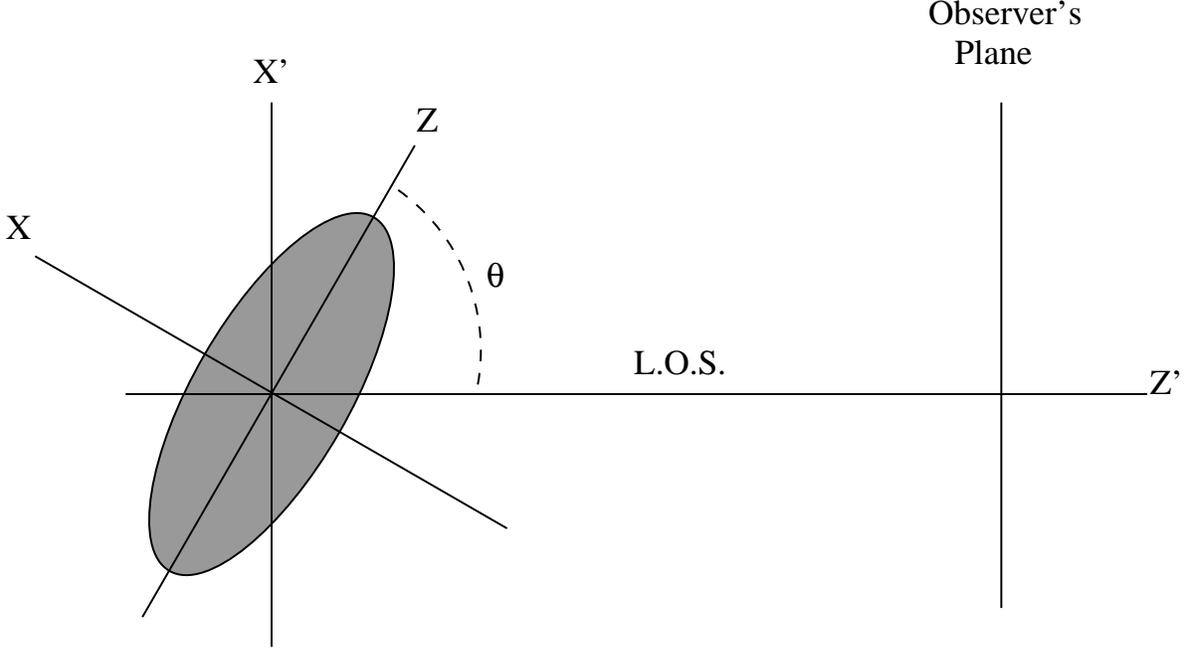}
\caption{ A diagram showing the coordinate systems used in the
paper. $(x^\prime,y^\prime,z^\prime)$ denote the the observers
coordinates with $z^\prime$ aligned with the line of sight. $(x,y,z)$
denote the cluster source function coordinates with $z$-axis is the cluster
symmetry axis.
}
\label{fig:config_digram}
\end{figure*}

The method for deprojection is described elsewhere
(\cite{zshs98}). Briefly, we deproject cluster images as follows. We
adopt the convention that bold-face symbols denote 3D quantities
(e.g., ${\bf k} = (k_x, k_y, k_z)$). Let the observer's coordinate
system be defined with the Cartesian axes
$(x^\prime,y^\prime,z^\prime)$, with the $z^\prime$ axis aligned with
the line of sight. Denote the (cluster) source function coordinate
system by the axes $(x,y,z)$ where the $z$-axis is the cluster
symmetry axis, forming an angle $\theta_i$ with respect to the line of
sight (see diagram shown in Figure \ref{fig:config_digram}).  Let
$I(x^\prime,y^\prime)=\int\lambda^\prime({\bf x^\prime})dz^\prime$
denote a projected quantity (image) of the source function
$\lambda$. The 3D Fourier transform (FT) of the source function is
related to the image Fourier transform by
\begin{eqnarray}
\lambda_{\bf k^\prime}^\prime(k_x^\prime,k_y^\prime,0)   
  &  = &\int e^{[-i(k^\prime_x x^\prime + k^\prime_y y^\prime)]}
I(x^\prime,y^\prime) \, {\rm d} x^\prime \, {\rm d} y^\prime  \nonumber \\
  & =  & I_{k^\prime}(k_x^\prime,k_y^\prime).  \label{eqn:simagereln} 
\end{eqnarray}

The expression relating the FT of the source
function in the observer and cluster rest frames is obtained simple by
coordinate rotation, so that
\begin{eqnarray}
                   \lambda_{k^\prime}^\prime({\bf k^\prime})&  = &
                   \lambda_k  \biggl( \sqrt{(-k_z^\prime \sin \theta_i +
k_x^\prime
                   \cos \theta_ i)^2
                  + k_y^{\prime\, 2}}, k_z^\prime \cos \theta_i + k_x^\prime \sin
                   \theta_i\biggr) \nonumber \\
                 & = &\lambda_{k}(k,k_z). \label{eqn:lambdareln}
\end{eqnarray}
where the last equality again holds due to axial symmetry. Inverse
transforming, we find the desired expression for the source
function in real space
\begin{equation}
\lambda^{\rm deproj}(r,z) = {\frac{1}{(2 \pi)^2}} \int  
	k\,dk\,dk_z \,e^{i k_z z}  J_0(k\, r) \,
I_{k^\prime}\biggl( \frac{k_z}{\sin \theta_i}, 
	\sqrt{k^2-k^2_z\cot^2 \theta_i}\biggr) . \label{eqn:maineqn}
\end{equation}

The implementation of this method is straightforward, assuming for the
moment that the inclination angle is known. The image FT is
evaluated at wavevectors  
\begin{equation}
k_x^\prime  =  k_z/\sin \theta_i  \;\;\; {\rm and} \;\;\; k_y^\prime  =
\sqrt{k^2 - k_z^2 \cot \theta_i^2} \label{eqn:wavevectors}
\end{equation}
and the inverse transform applied.

For wavevectors $k<\vert k_z\vert\,\cot\theta_i$, the argument of the
image FT becomes imaginary defining a cone in k-space known as the
``cone of ignorance'' (hereafter COI).  

An interesting property of equation (\ref{eqn:maineqn}) is that the
k-space information in the $k_z=0$ plane is fully known and independent
of the cluster inclination angle. It is, therefore, straightforward to
show, by integrating over Eq. 2 with respect to $z$, that the projection over the
$x'$-direction is related to the symmetry axis integrated density,
\begin{equation}
\rho^z(r)\equiv \int \lambda(r,z) dz = {\frac{1}{2 \pi}} \int k\,dk
J_0(k\, r)\,\nonumber I_{k^\prime}(0,k), \label{eqn:rhoz}
\end{equation}
where $I_{k^\prime}(0,k)$ is FT of $\int I(x^\prime,y^\prime)
dx^\prime$.  We can also define the z-projected cluster `mass',
$M^z(r)=\pi \int \rho^z(r)\,r\,dr$.

This relation is especially useful in the case of isothermal clusters,
where the SZ map is proportional to the projected electron number
density, hence to the projected gas density, and the surface density
map is proportional to the projected total mass density. Therefore
equation (\ref{eqn:rhoz}) provides an angle independent estimate of
$\Omega_b/\Omega$, and enables evaluation of the ratio $\rho^z_{\rm
gas}/\rho^z_{DM}$ and $M^z_{\rm gas}/M^z_{DM}$ as a function of the
distance from the symmetry axis.  
Note also that since the gas and total mass distributions for an
isothermal cluster are roughly proportional to each other in most of
the volume (\cite{ettori99}), except the inner most region, this
comparison is almost Hubble constant independent.

\section{The Cluster Sample} \label{sec:the_simulations}

To test the validity of the deprojection method, we employed a set of
outputs from a gas dynamic simulation of the formation of an X-ray
cluster.  The simulation is Evrard's contribution to the Santa Barbara
cluster comparison project (\cite{frenk98}). Details of the
simulation are provided elsewhere (\cite{frenk98}), but we very
briefly outline the data set here:

The underlying structure formation model was CDM, with the G3 transfer
function fit of Bardeen \etal (1986) at $z=20$. The simulation was
performed in a flat universe, the initial fluctuation spectrum was given
a shape parameter of $\Gamma = 0.25$ and normalized such that the
present day rms mass fluctuations in a spherical top-hat of radius $16\,
\Mpc$ is $\sigma_8 = 0.65$. The cosmological parameters were assigned
with as $H_0 = 50 \; {\rm km/s/Mpc}$, $\Omega = 1$, and a contribution
from the baryons of $\Omega_b = 0.1$.

The initial conditions for the simulation were generated with a
constrained Gaussian random field (\cite{hoffman91}), and galaxy
clusters identified with $3 \sigma$ peaks of the field smoothed with a
Gaussian filter of scale $10 \, \Mpc$, and centered in a cubic region of
side $64 \,\Mpc$.

The evolution of the gas was followed using the Smoothed Particle
Hydrodynamics (SPH) technique (\cite{evrard88}) to $z=0$.  Final maps of
the projected total mass distribution, Compton y-parameter, and
simplified versions of the X-ray luminosity and X-ray emission weighted
temperature were produced. The projected quantities were calculated
from a $(32\, \Mpc)^3$ cube, and computed on a $256 \times 256$ grid,
corresponding to a physical scale of $10\, \Mpc$ at $z=0$. In the
Rayleigh-Jeans regime, the Compton y-parameter is related to the SZ
decrement as $\frac{\delta T} { T} = -2 y$. We have applied this
conversion to the simulation output.

For the purposes of this test, we employed three orthogonal projections
of a single cluster at four redshifts: $z = 0, 0.3, 0.6,
\;{\rm and} \; 0.9$. This, in effect, yields a sample of 12
semi-independent clusters to study with a wide range of dynamical states
and physical distributions of the gas and dark matter. For example, at
$z = 0.6$, the cluster undergoes a major merger event, and thus the
outputs at that redshift permit a study of the method while the cluster
is in a highly unrelaxed state.

\section{Application: The Prototype Cluster}

\begin{figure*}
\plotone{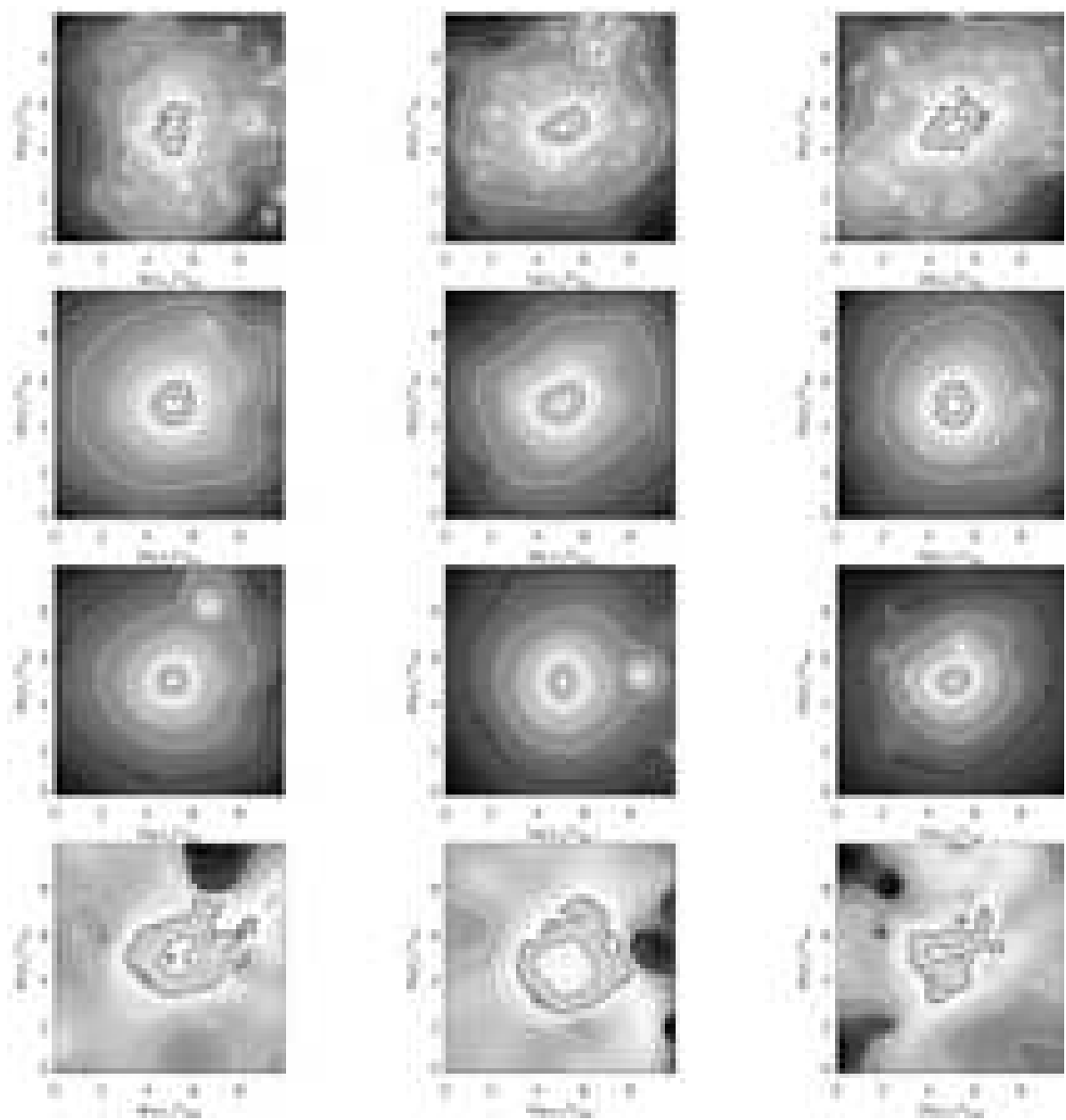}
\caption{Left column: The x-projection for the prototype cluster. The
grey scale and contours show the logarithmic intensity. Surface density
(first panel; $M_\odot/{\rm Mpc}^2$), and Sunyaev-Zel'dovich decrement
(second panel). X-ray surface brightness (third panel; $M_\odot^2
K^{1/2} /{\rm Mpc}^5$), and emission weighted temperature (fourth panel;
$K$). The middle and right columns show the same plots for the y- and
z-projections respectively.}
\label{fig:allproj}
\end{figure*}

For the first part of this analysis, we concentrate on a single
`prototype cluster'; the simulation output at $z = 0$. In Figure
\ref{fig:allproj}, we display the logarithmically scaled mass surface
density, X-ray surface brightness, SZ decrement and emission weighted
temperature distributions for the three orthogonal projections of the
prototype cluster. The dimensions for the surface density, temperature
and SZ decrement are physical ($M_\odot/{\rm Mpc}^2$ and $K$ for the
first two, while the SZ decrement is dimensionless). The X-ray surface
brightness was calculated as the line of sight integral $\int \rho_{\rm
gas}^2 T^{1/2} dl$ and has dimensions $M_\odot^2 K^{1/2} /{\rm
Mpc}^5$. In this section, unless stated otherwise, the images used for
deprojection are the y-projection maps of the simulated protocluster.

For the prototype cluster, the projected gas and total mass
distributions are elliptical, with axis ratios $a/b \gtrsim 2:1$. In all
of the projections, an infalling sublump is evident, approximately
$\simeq 3 \, \Mpc$ (projected distance) from the cluster center. We note
that the emission weighted temperature is nearly constant over the
innermost $\simeq 3\, \Mpc$, with a value of $(7.5\pm
0.5)\times10^7\,K$. In what follows for this specific cluster, we treat
the gas distribution as being isothermal.

\subsection{The Baryon Fraction}

\begin{figure*}[htbn]
\plottwo{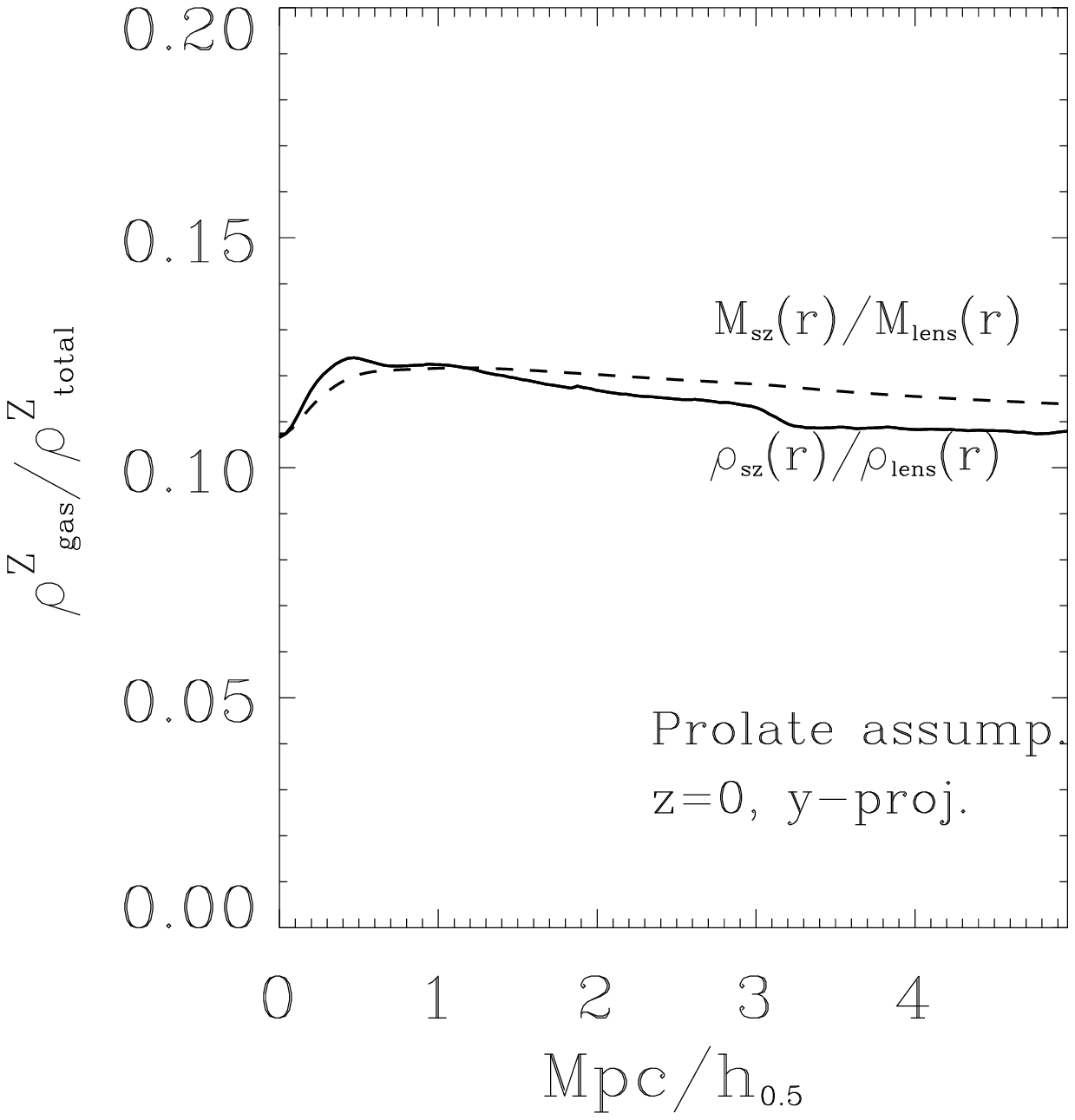}{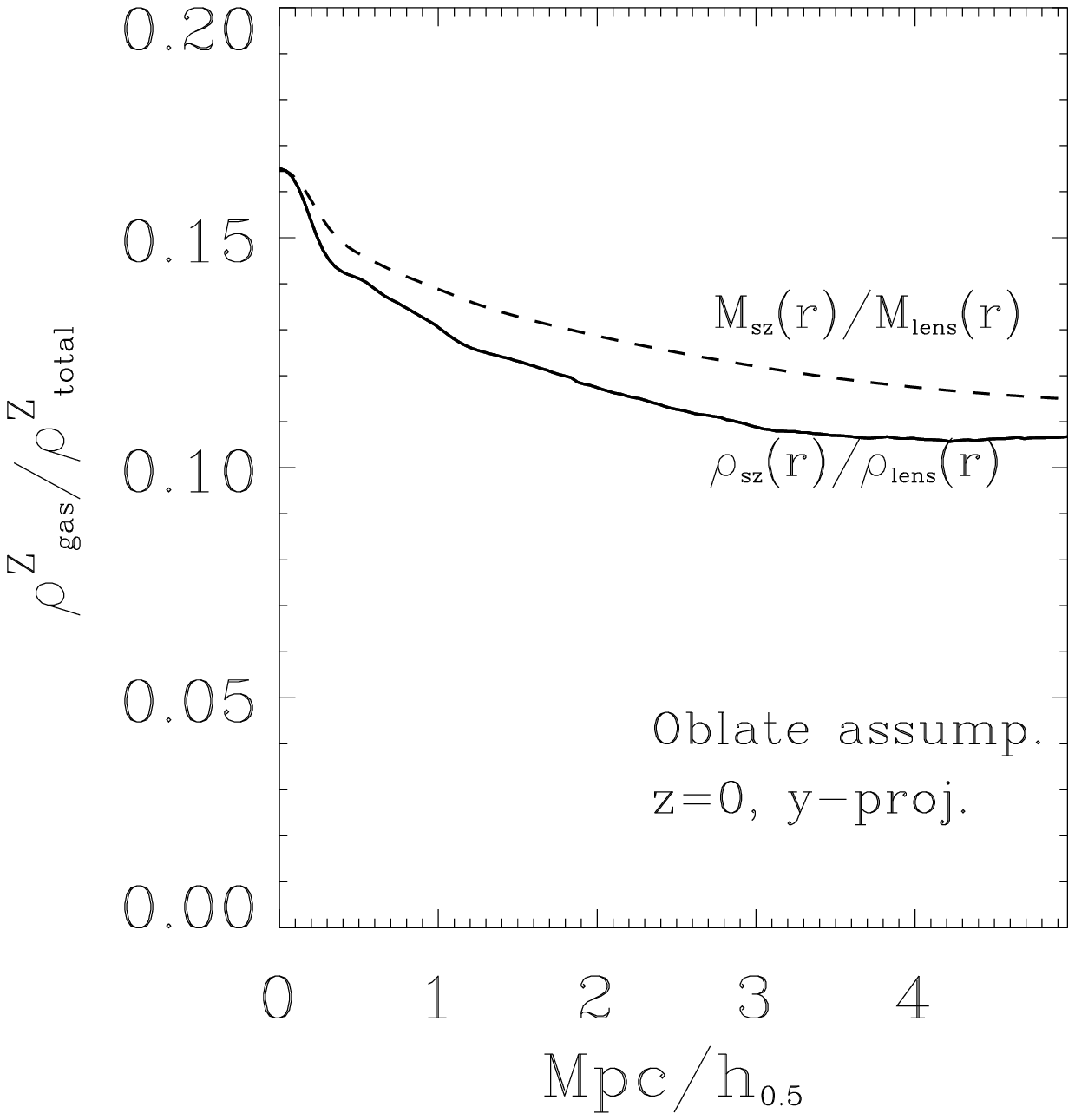}
\caption{The ratio of the z-integrated gas density and dark matter
density, $\rho^z_{gas}/\rho^z_{DM}$ , as a function of r, calculated
from the surface density and Sunyaev-Zel'dovich maps of the y-projection
shown in the middle row of Figure \ref{fig:allproj} (solid line) and
the corresponding gas to dark matter mass ratio (dashed line).  The left
and right panels are produced assuming a prolate and oblate cluster
respectively.}
\label{fig:densityratio}
\end{figure*}

Using equation (\ref{eqn:rhoz}), we calculated the gas to total mass and
density ratios for prototype cluster. The results are shown in Figure
\ref{fig:densityratio} for the y-projection maps.  As discussed below
(see \S \ref{sec:determining_projected_symmetry_axis}), we do not know
apriori if the cluster is prolate or oblate, and so we have calculated
the ratios under both assumptions. However, this is the only unknown; no
knowledge of the inclination angle is required to determine these
ratios.

The value of density and mass ratios in the case of this prototype
cluster is $\sim 0.1$. Under the assumption the cluster is prolate, this
ratio is approximately constant at all radii, whereas in the oblate
case, it rises by $\simeq 50$\% towards the center. However, in both
cases, at the outermost radii probed by our calculation, the density
ratio becomes $\sim 0.105$ which is extremely close to the input ratio
in the SPH/N-body simulation.

The density and mass ratios across the cluster, especially in relaxed
clusters, are expected to have an almost constant value, with small
radial dependence (\cite{ettori99}). Therefore, the greater
variability shown in the density and mass ratios under the assumption
of oblate cluster might be used as an indicator to that the
protocluster at hand is most probably prolate.  Inspection of the true
3D distribution shows that this is indeed the case. The same argument
can be repeated for all the clusters we have analyzed in this paper
(see Figs.~\ref{fig:z00_densityratio}, ~\ref{fig:z03_densityratio},
~\ref{fig:z06_densityratio}, and ~\ref{fig:z09_densityratio}).

Furthermore, the true full 3D distribution clearly shows that the gas
radial density profile follows very faithfully the radial mass density
profile down to the inner most $0.3 {\rm Mpc}$, interior to this
radius the ratio drops down due to the effect of the pressure (see
Figs. 10 and 12 in \cite{frenk98}), which explains the drop in the gas
to mass ratio at the inner most radii in the prolate case shown in the
left panel of Figure~\ref{fig:densityratio}.

\subsection{The Cluster 3D Structure}

To demonstrate the utility of the method for the determining the
underlying 3D structure, we proceed in two ways: first, we demonstrate
that a robust and stable determination of the inclination angle can be
achieved using the sort of information that would be available from
observations (i.e., X-ray, SZ and lensing maps). Secondly, using our
knowledge of the true underlying 3D structure, available since we are
employing simulated clusters, we determine the accuracy with which
cumulated quantities (such as gas or total mass) can be determined
from this method.

We encountered several subtle features of the deprojection procedure
which we describe in detail below. However, the overall procedure is,
conceptually at least, straightforward: we first determine the image
axis of symmetry, which is the projection of the axis of symmetry in the
3D distribution. We allow both possibilities that the cluster is either
oblate or prolate, and deproject using equation (\ref{eqn:maineqn}).  As
the inclination angle is unknown, we deproject each image over a wide
range of assumed inclination angles, searching for the angle that
consistently yields the best reconstructed ratios between the
Sunyaev-Zel'dovich, X-ray and the total mass deprojections, as discussed
below.

\subsubsection{Determining the Symmetry Axis from Deprojection of the Images}
\label{sec:determining_projected_symmetry_axis}

To perform the image deprojection, the first step is to define the
(projected) symmetry axis. This turns out to have a couple of subtle
complications: First, the simulation data often seems to have, at least
qualitatively, a varying axis of symmetry as a function of radius from
the image centroid. Secondly, the images become progressively rounder
from the surface density, to SZ, to X-ray maps respectively -- how do
the axes of symmetry found from each independently compare?

To determine the projected symmetry axis, we proceed as follows.  Denote
the image by $f[x,y]$, where $ x, y \in [0, N-1]$, and $N$ is the image
size ($N=256$ for this simulation data).  Suppose the axis of symmetry
is parallel to the $y$-axis, passing through the point $x = x_c$. Then,
for a perfectly axially symmetric object
\begin{equation}
f[x_c - x, y] = f[x_c + x, y] \;\;\; \forall \, x,y. 
\label{eqn:axiallysymmetry}
\end{equation}
Given a real image (which may only be approximately axially symmetric),
the axis of symmetry is the one that, in some sense, `best' satisfies
equation (\ref{eqn:axiallysymmetry}). To determine this axis
empirically, we scan through a set of points around $(x_c, y_c)$, (the
starting point is usually chosen to be the coordinates of the object's
`center of mass') and find the line that best satisfies the axially
symmetric condition as quantified below.

In the case where the pixel counts are uncorrelated, and the model is
that the image is axisymmetric with Poisson noise, then we form a
$\chi2$-like statistic that is minimized when the image is rotated with
the axis of symmetry aligned with the y-axis:
\begin{equation}
R_1(\theta) = \frac{1}{N_{pix}} \sum_{x,y}  \left(f_{\rm rot}[x_{c_n} - x, y] - 
        f_{\rm rot}[x_{c_n} +x, y] \right)^2. \label{eqn:rsum}
\end{equation}

One consideration is that a $\chi^2$ estimator is only strictly valid in
the limit that the errors follow a Gaussian distribution. 
However, in practise, X-ray observations are counting experiments, often
with low signal-to-noise, and such estimators are not valid. An
alternative is to consider a robust M-estimator (\cite{press92}). If we
assume that the that the errors are distributed exponentially, then the
maximum likelihood estimator for the axis angle is determined by
minimizing over the sum
\begin{equation}
R_2(\theta) = \frac{1}{N_{pix}} \sum_{x,y}  
        \left|\, f_{\rm rot}[x_{c_n} - x, y] - 
        f_{\rm rot}[x_{c_n} +x, y] \, \right|^{1/2} . \label{eqn:msum}
\end{equation}

We also considered an alternative, and less formal, approach to
determining the symmetry axis. This was partially motivated by the
shortcomings of the $\chi^2$ estimator for Poisson errors. Furthermore,
since the X-ray and, to a lesser extent, the radio emission and mass
density, are typically quite centrally concentrated, in the Poisson
model for the noise, there is little contribution to the $R_1$ and $R_2$
sums from data at large radii from the cluster center, even if the
difference in counts between symmetric pixels is large. Thus, the above
estimators will be dominated by terms only from the central region of
the cluster. To enable us to be sensitive to isophote rotation, we have
adopted relative difference between symmetric pixels, which will give
equal weight to {\em all} image pixels
\begin{equation}
R_3(\theta) = \frac{1}{N_{pix}} \sum_{x,y}  
        \left| \frac{f_{\rm rot}[x_{c_n} - x, y] - 
        f_{\rm rot}[x_{c_n} +x, y]}{f_{\rm rot}[x_{c_n} - x, y] + 
        f_{\rm rot}[x_{c_n} +x, y]}  \right|. 
\end{equation}
A `good fit' for the symmetry axis will have $R_3(\theta) \rightarrow
0$, while poor fits will have $R_3(\theta) \rightarrow 1$.

We originally were interested to see if we could find an optimal
estimator of the symmetry axis, based on equation
(\ref{eqn:axiallysymmetry}). However, the brute force implementation of
the sums to compute $R_1$, $R_2$ and $R_3$ are computationally very
fast, and hence we have not pursued this further. In calculating all of
these statistics to find the image axis of symmetry, we have allowed the
point $(x_c, y_c)$ to be a free parameter in the fit. This becomes a
highly non-linear minimization problem, and we have used the simplex
method (\cite{nelder65}) to simultaneously determine the axis angle and
centroid.

In what follows, we have employed the $R_3$ to determine the projected
image axis. This choice was made somewhat arbitrarily; our rationale was
that since we are primarily interested in determining the long
wavelength cluster features, we prefer to give increased weight to the
image counts at large radii from the cluster center. In practice, this
choice seems to work very well for recovering the cluster 3D structure,
as we demonstrate in what follows below. However, we do note that using
either the $R_1$ or $R_2$ statistics also give good results, and we do
not have a definitive recommendation as to which to use in analyses of
real data (in fact, during our preliminary work with the simulated
cluster sample for this study, we used all three estimators and a
similar procedure would be prudent when applying the method to real
data).

One final complication arises in that apriori we do not know if the
cluster is prolate or oblate. That is, in general, there will be two
axes that best satisfy the axial symmetry condition corresponding to the
case that the cluster is oblate or prolate. In the context of the
deprojection and what follows, we use both axes and test via a
comparison of the resulting 3D structure under each hypothesis which is
valid.

\subsubsection{Filling the COI}
\label{sec:COI_filling}

Having defined the image axis of symmetry, the next operation in the
deprojection algorithm requires filling the COI of each deprojection. In
paper I, a simple extrapolation scheme was used to fill the
COI. Here we have explored somewhat more sophisticated extrapolation
methods, such as fitting an elliptical isothermal model to the actual FT
of the image outside the COI, and using the model fit to extrapolate
into the COI. We find that our results are fairly insensitive to the
precise algorithm applied to fill the COI, primarily because the
information lost inside the COI is restricted to moderately high spatial
frequencies. Whatever scheme we apply to fill the COI affects most the
small scale details of the 3D densities we infer, while the large scale
features are less affected.

\begin{figure*}[htbn]
\plotthree{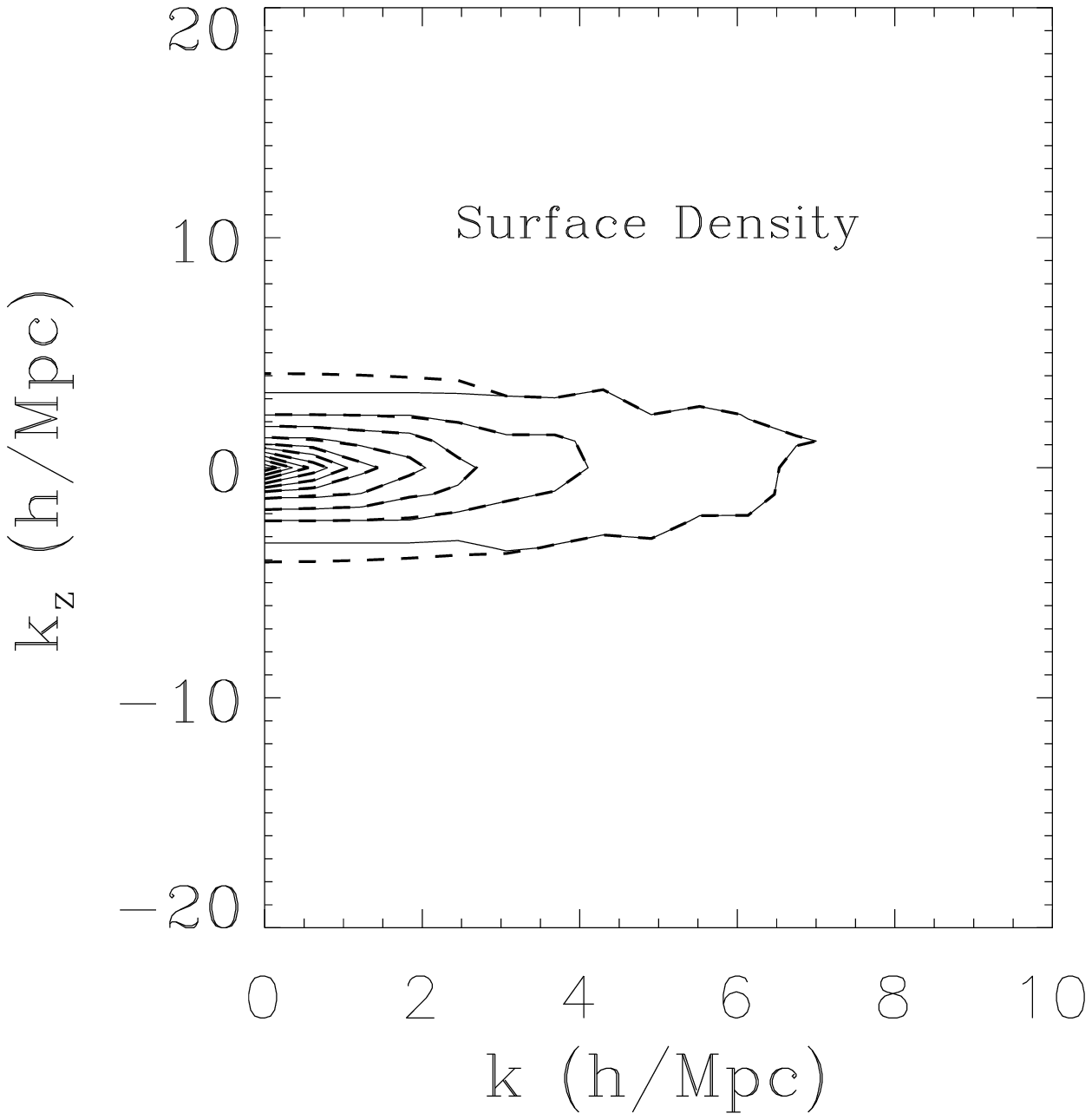}{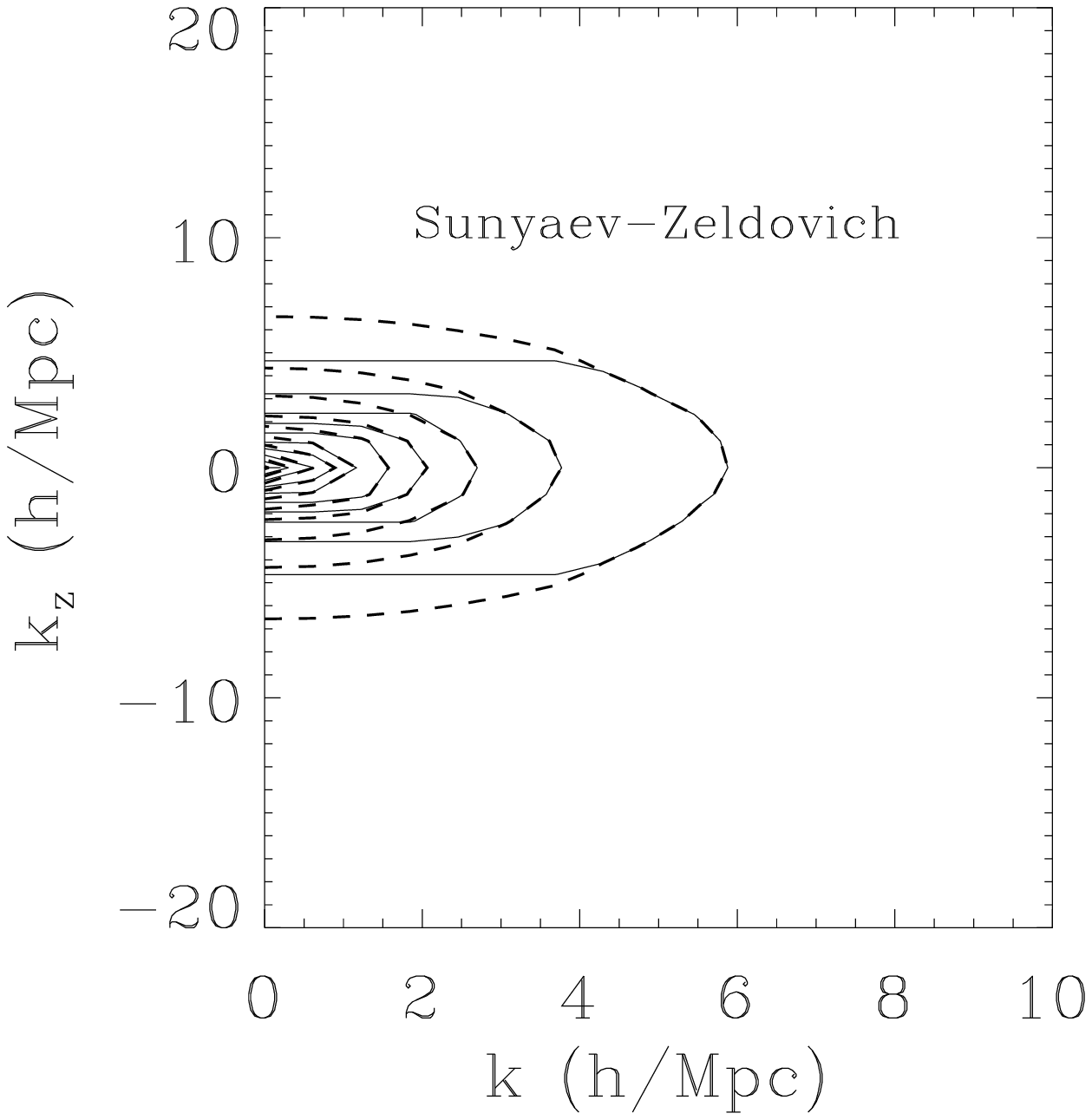}{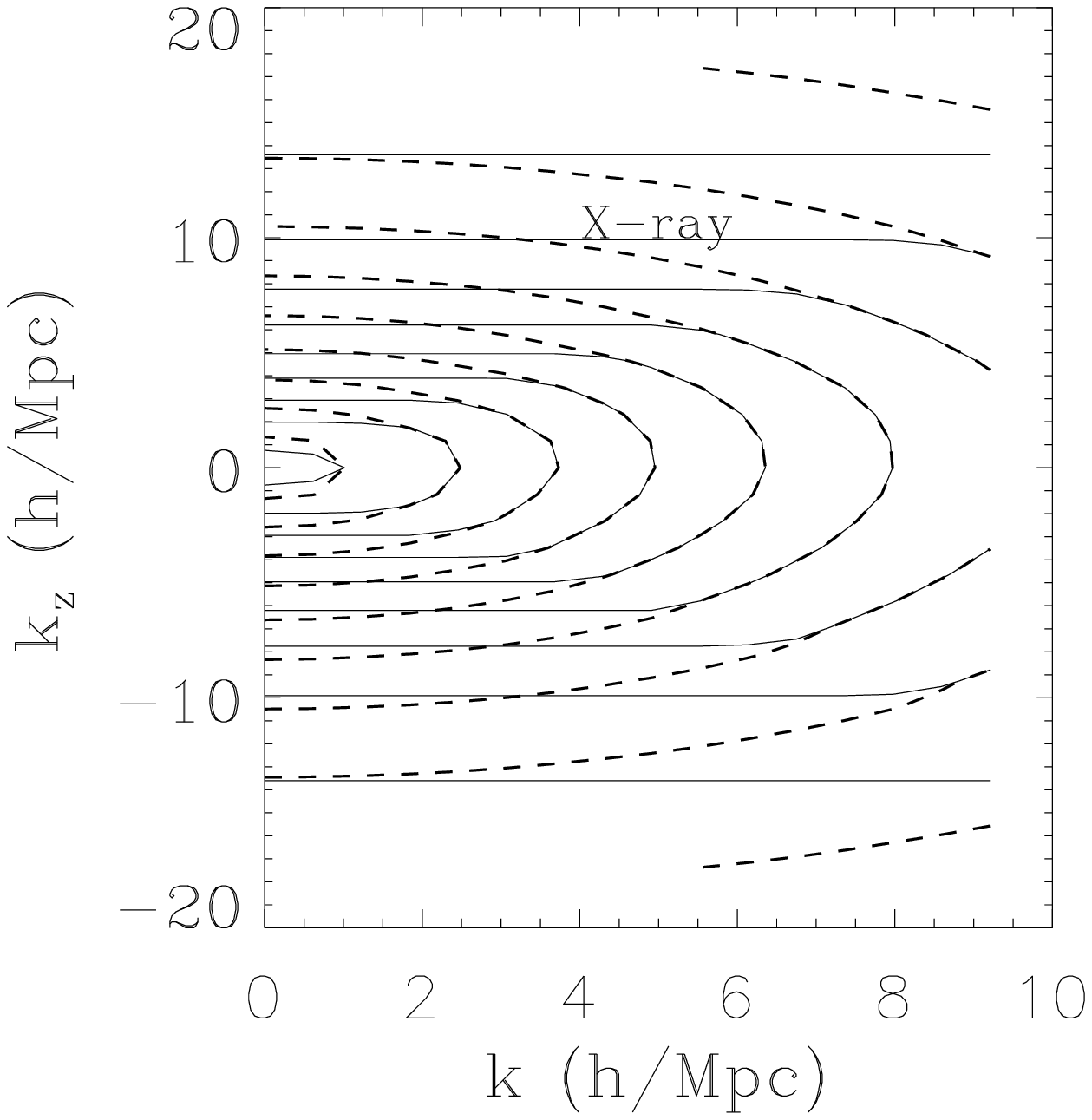}
\caption{The structure of the source function in k-space is shown for
the surface density, Sunyaev-Zel'dovich and X-ray maps. The COI is
filled with a simple linear extrapolation (solid line), and an
isothermal elliptical profile (dashed line). For this example, the
cluster is assumed to be prolate with $32^\circ$ inclination angle.}
\label{fig:coi}
\end{figure*}

This is clearly seen in Figure \ref{fig:coi} where we plot the k-space
structure of the SZ, X-ray and surface density maps, where we have
assumed for the moment that the inclination angle is $32^\circ$ (this
simply defines the region occupied by the COI. We motivate this choice
of angle in \S \ref{sec:determining_theta}).  Most of the long
wavelength power is outside the COI, which tends to dominate k-space
primarily at fairly high spatial frequencies.

Still, in order to reduce the effect of ringing by the discontinuities
at the edge of the COI, we need to do some sort of extrapolation into
this regime. We display two alternatives here: first we simply perform a
linear extrapolation into the COI with the amplitude fixed by the value
at the cone boundary. This is shown by the solid line in Figure
\ref{fig:coi}. Secondly, we note that for this cluster, the image FT of
the SZ and X-ray images are reasonably well fit by the elliptical
isothermal gas density model, and we have used this model to extrapolate
the observed image FT into the COI with the dashed line in Figure
\ref{fig:coi}. This prescription seems quite reasonable for the
Sunyaev-Zel'dovich and X-ray surface brightness maps, while the fit for
the surface density map is not as successful. This is somewhat expected
as the former (which probe the gas density distribution) are smoother
and not as sensitive to the details manifest in the cluster's
substructure as the surface density map.

\subsubsection{Determining the Inclination Angle and Image
Deprojection} 
\label{sec:determining_theta}

Having a prescription to fill the COI, we are now in position to do the
deprojection, compare the results from the various projections, and
determine the cluster inclination angle.  We first consider the
comparison of the SZ {\it vs.} X-ray deprojections. We deproject each
image independently (with the same inclination angle), solve for the 3D
gas density, and compare the results from each deprojection.

\begin{figure*}[htbn]
\plottwo{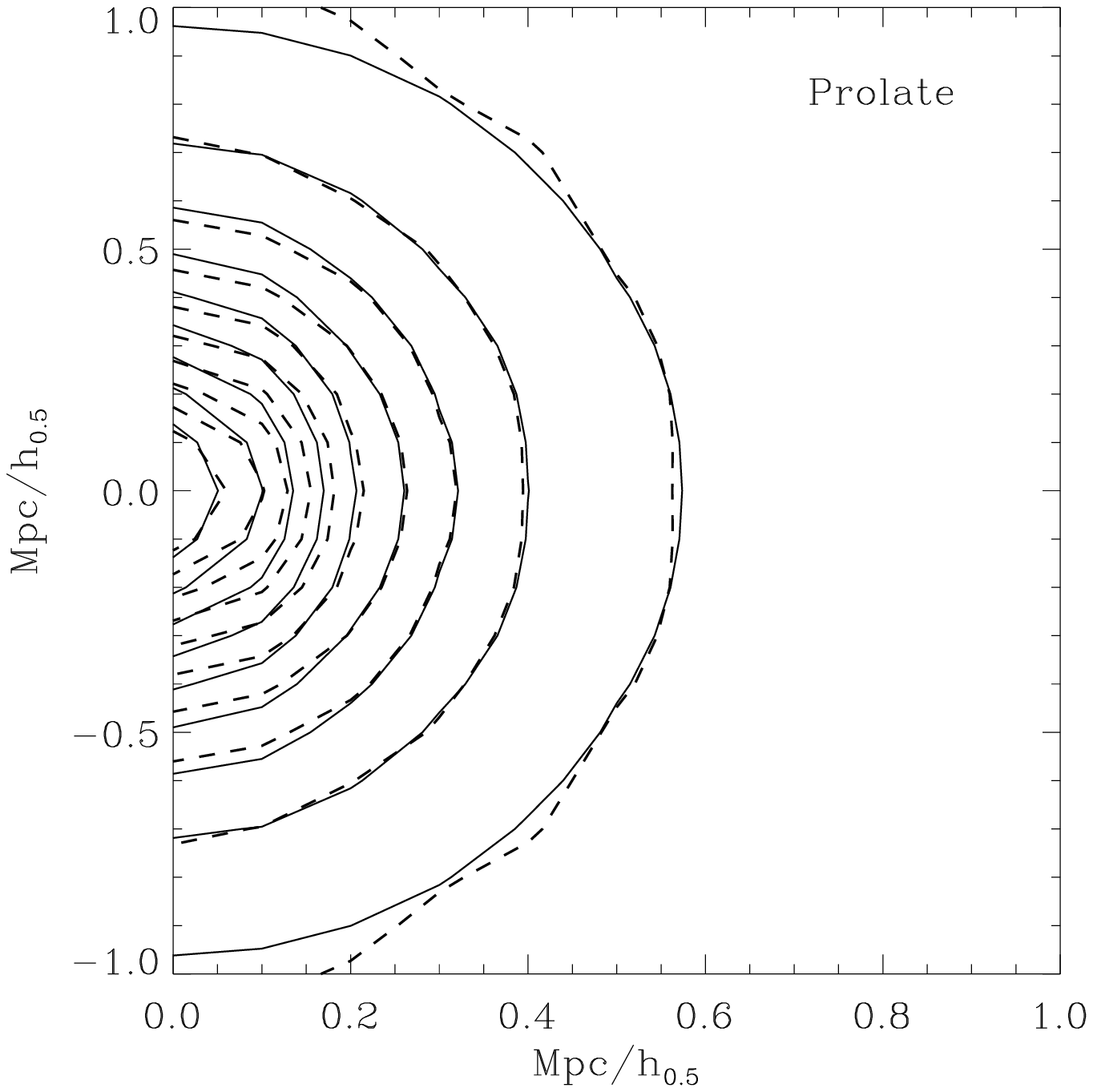}{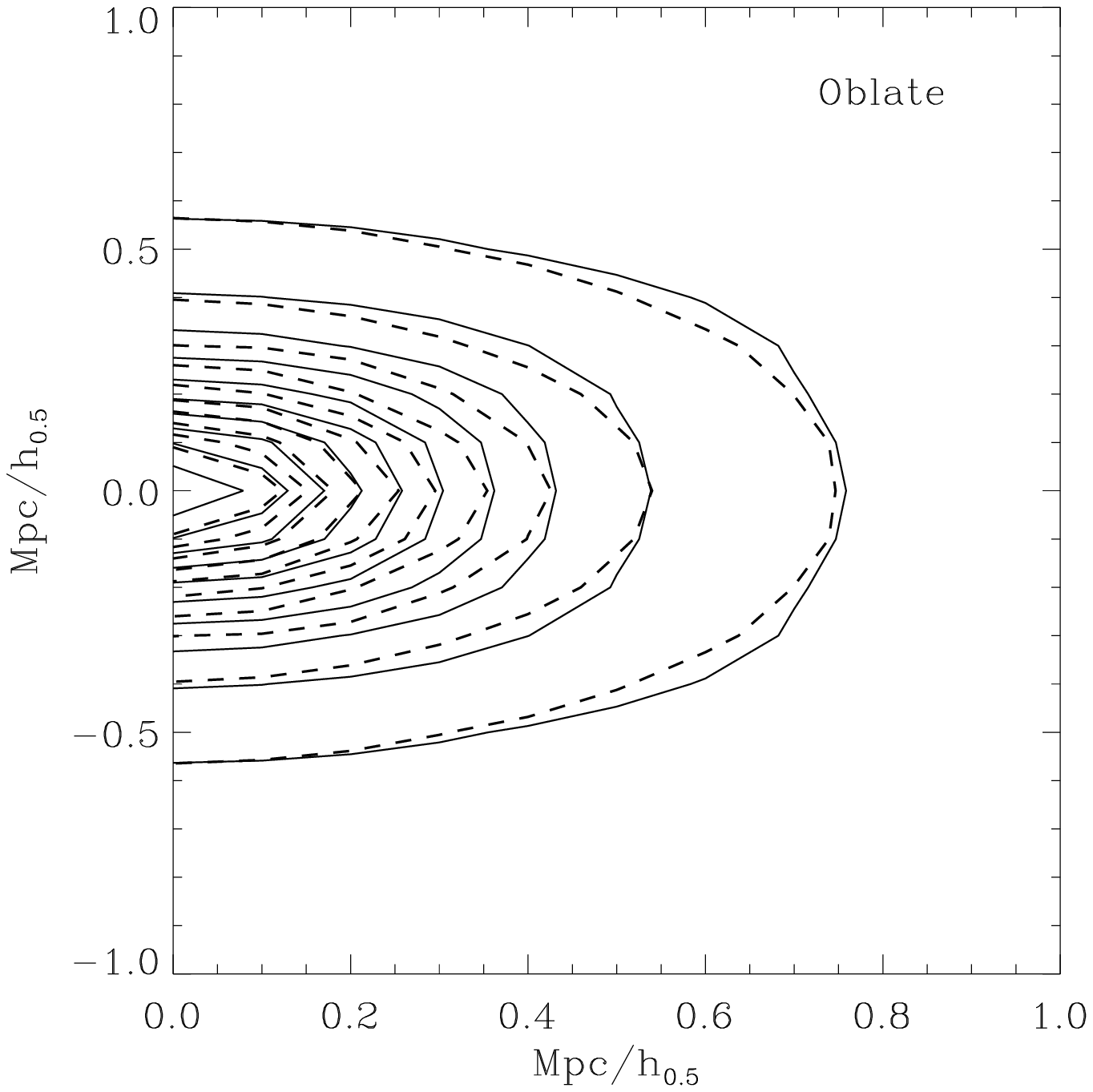}
\caption{The deprojected gas density distribution from the SZ image
(solid lines) and from the deprojection of the X-ray image (dashed
lines). The prolate case is shown in the left panel
($\theta_i=32^\circ$), while the oblate case is shown in the right
panel ($\theta_i=65^\circ$).}
\label{fig:SZvsXray}
\end{figure*}

 In paper I, we demonstrated that in order to determine the inclination
angle from a comparison of the Sunyaev-Zel'dovich and X-ray maps, the
{\em shape} alone of the density profiles from each image is
insufficient as the method, for any given angle, produces the same
profile shape from both maps. The only way to determine the angle is
therefore from the ratio of the {\em amplitudes}. This ratio is used
to determine the inclination angle (where we use our knowledge, for the
moment, that $h=0.5$) yielding a best fit value of $\theta_i=32^\circ$
and $\theta_i=65^\circ$ for prolate and oblate cluster respectively.
Figure~\ref{fig:SZvsXray} shows the gas density distribution as inferred
from the deprojection of the SZ and X-ray maps for prolate and oblate
clusters using the appropriate best fit angle for each case. The
agreement between the two, in terms of both shape and amplitude, is very
good (within $5-10\%$ difference at the center).

The comparison between the Sunyaev-Zel'dovich and surface density maps
is somewhat more complicated, as we need to apply the hydrostatic
equation to relate the dark matter and gas distributions
\begin{equation}
 \nabla({\nabla(\rho_gk T_x/\mu m_p)\over \rho_g})=-4\pi G \rho_{tot},
\label{eqn:hydro}
\end{equation}
where $\rho_g$, $\rho_{tot}$ and $T_x$ are the gas and total mass
densities and the cluster's temperature respectively.  The gas is
assumed to be an ideal gas. For an isothermal cluster this equation
becomes very simple, $\nabla^2\ln\rho_g = \alpha\rho_{tot}$, where
$\alpha$ is a (known) numerical constant. Note that in the limit of
$\Omega_b/\Omega \ll 1$ equation (\ref{eqn:hydro}) is independent of the
amplitude of the gas density, $\rho_g$ and therefore is sensitive to the
shape of the gas profile. This forms an attractive supplement to the SZ
versus X-ray comparison, which depends only on the amplitude.

\begin{figure*}[htbn]
\plottwo{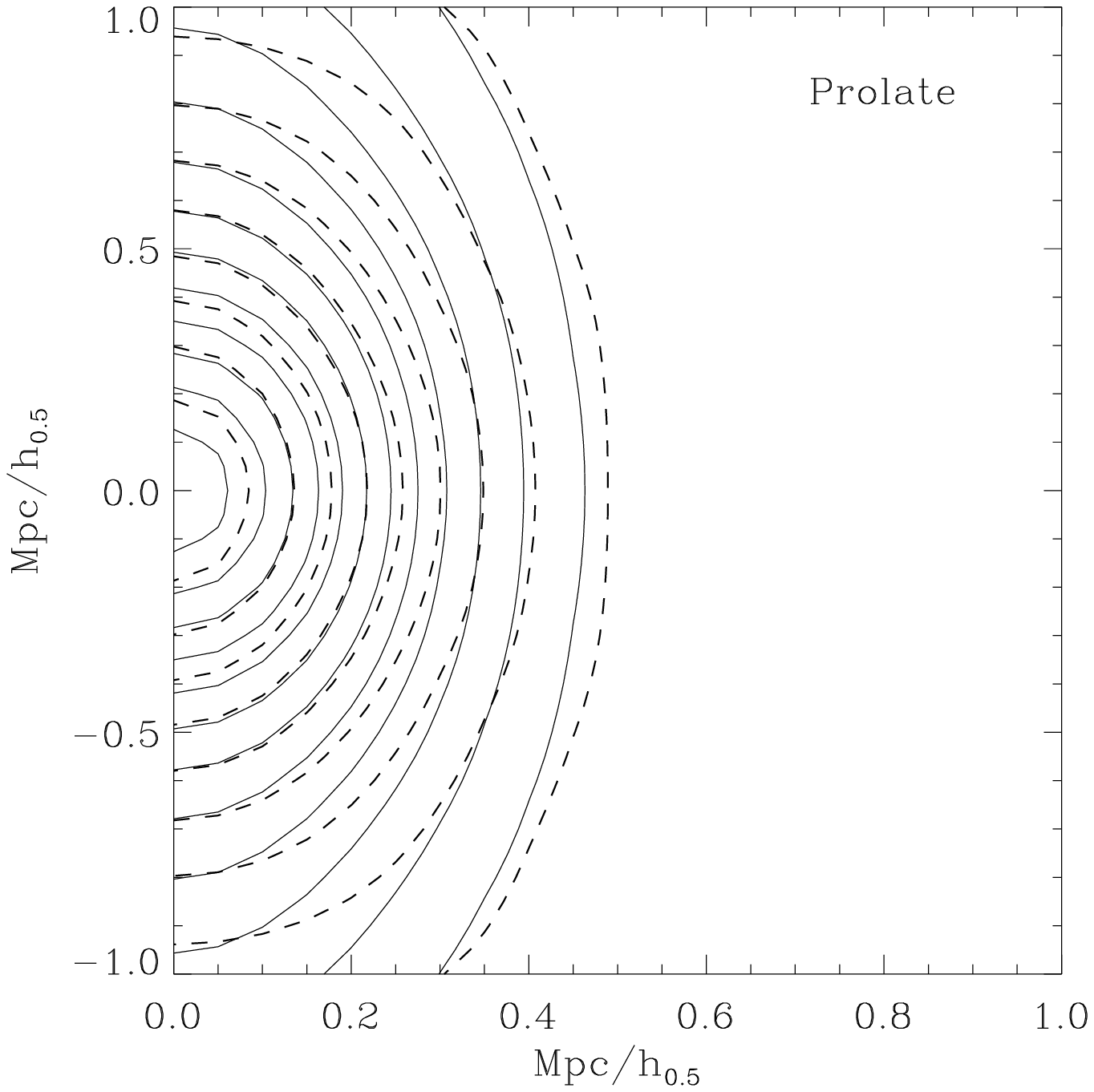}{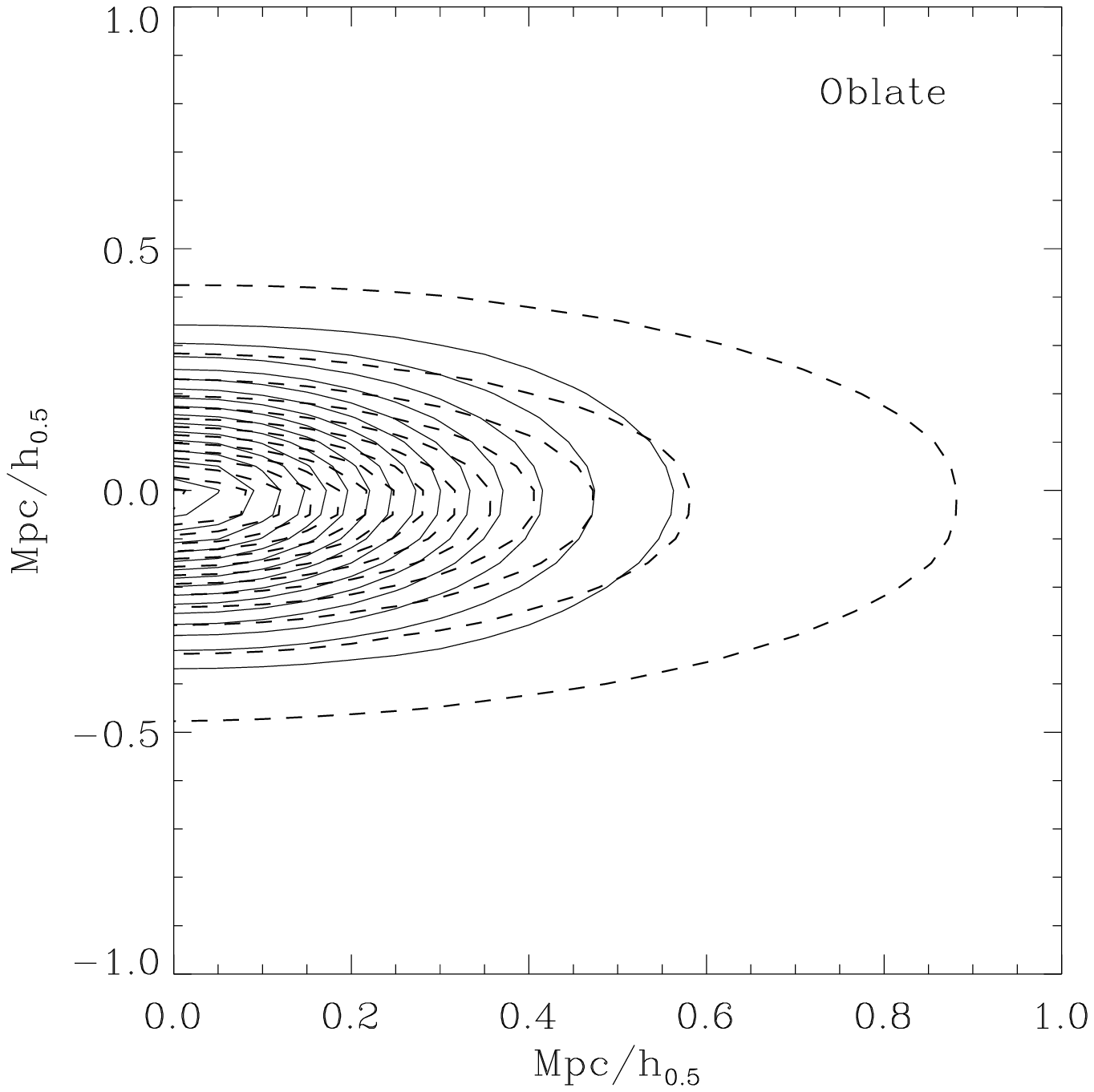}
\caption{The deprojected dark matter density distribution from the
surface density image (solid lines) and from, via the hydrostatic
equation, the deprojection of the SZ image (dashed lines). The prolate
case is shown in the left panel ($\theta_i=32^\circ$), while the oblate
case is shown in the right panel ($\theta_i=65^\circ$).}
\label{fig:DMvsgas}
\end{figure*}

Operationally, to compare the SZ and surface density deprojections we
proceed in a similar fashion as with the SZ and X-ray maps: the SZ and
surface density maps are deprojected for many angles assuming oblate and
prolate structures. For each angle the gas density is inserted in
equation (\ref{eqn:hydro}) and the result is compared with the total
density profile by a $\chi^2$-like statistic.  The best fit between the
two again occurs at the angles $32^\circ$ and $65^\circ$ for the prolate
and oblate assumptions respectively. These comparisons are shown in
Figure \ref{fig:DMvsgas}.

\begin{figure*}
\plotone{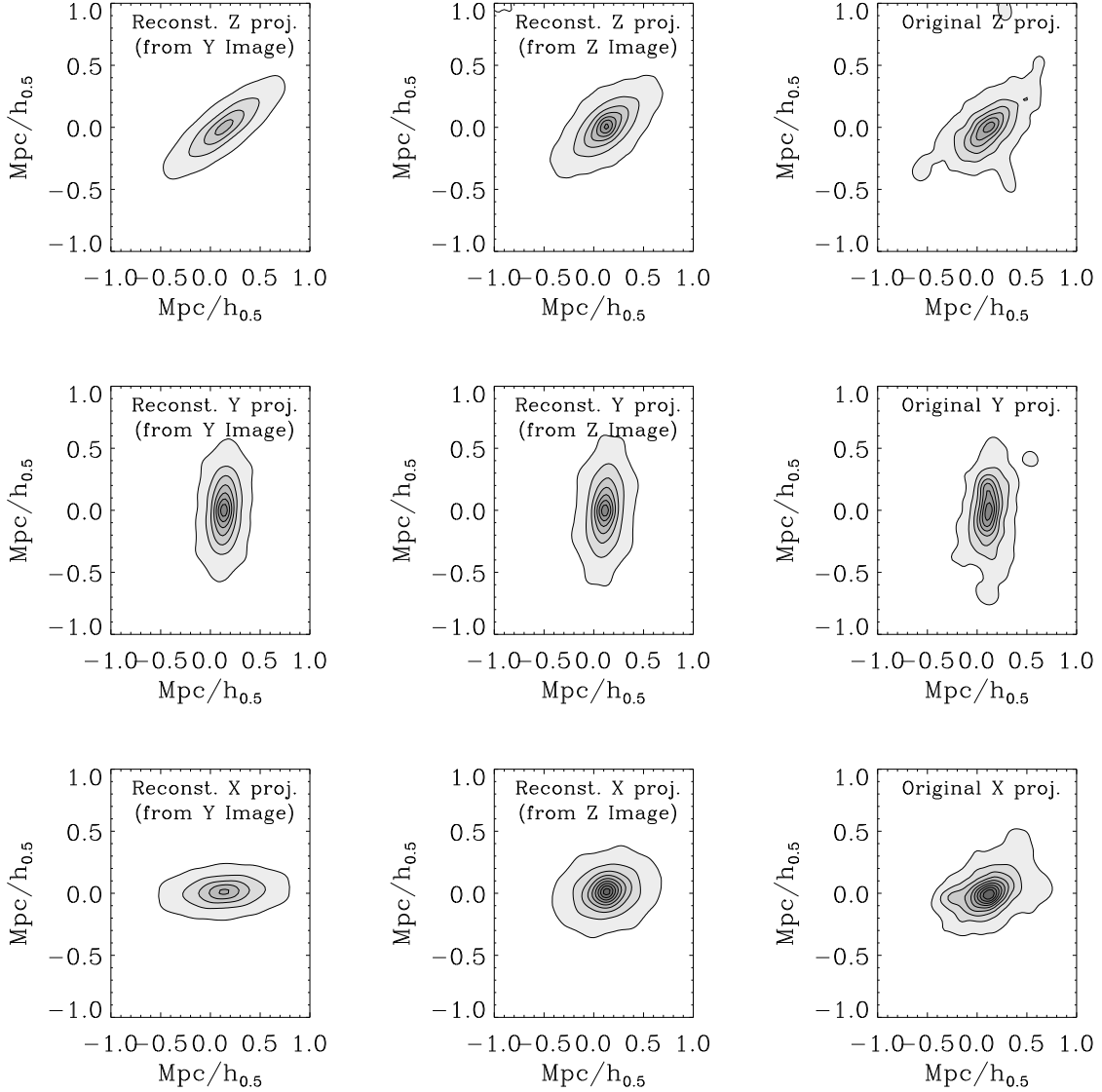}
\caption{We determine the inclination angle by requiring consistency in
the deprojection of the y-projection of the X-ray and SZ images. Using
this, we {\em predict} the orthogonal projections of the surface density
maps and compare with the true distributions. The left panels display
the predicted x- y- and z- projections where the input image was taken
to be the `observed' y-projection SZ and X-ray images (and
$\theta_i=32.^{\circ}$). The middle panel shows the result where the
input image was taken to be the `observed' z-projection SZ and X-ray
images (and $\theta_i=84.^{\circ}$). The right panel shows the true
underlying total mass distribution.}
\label{fig:projection_lens}
\end{figure*}

\begin{figure*}
\plotone{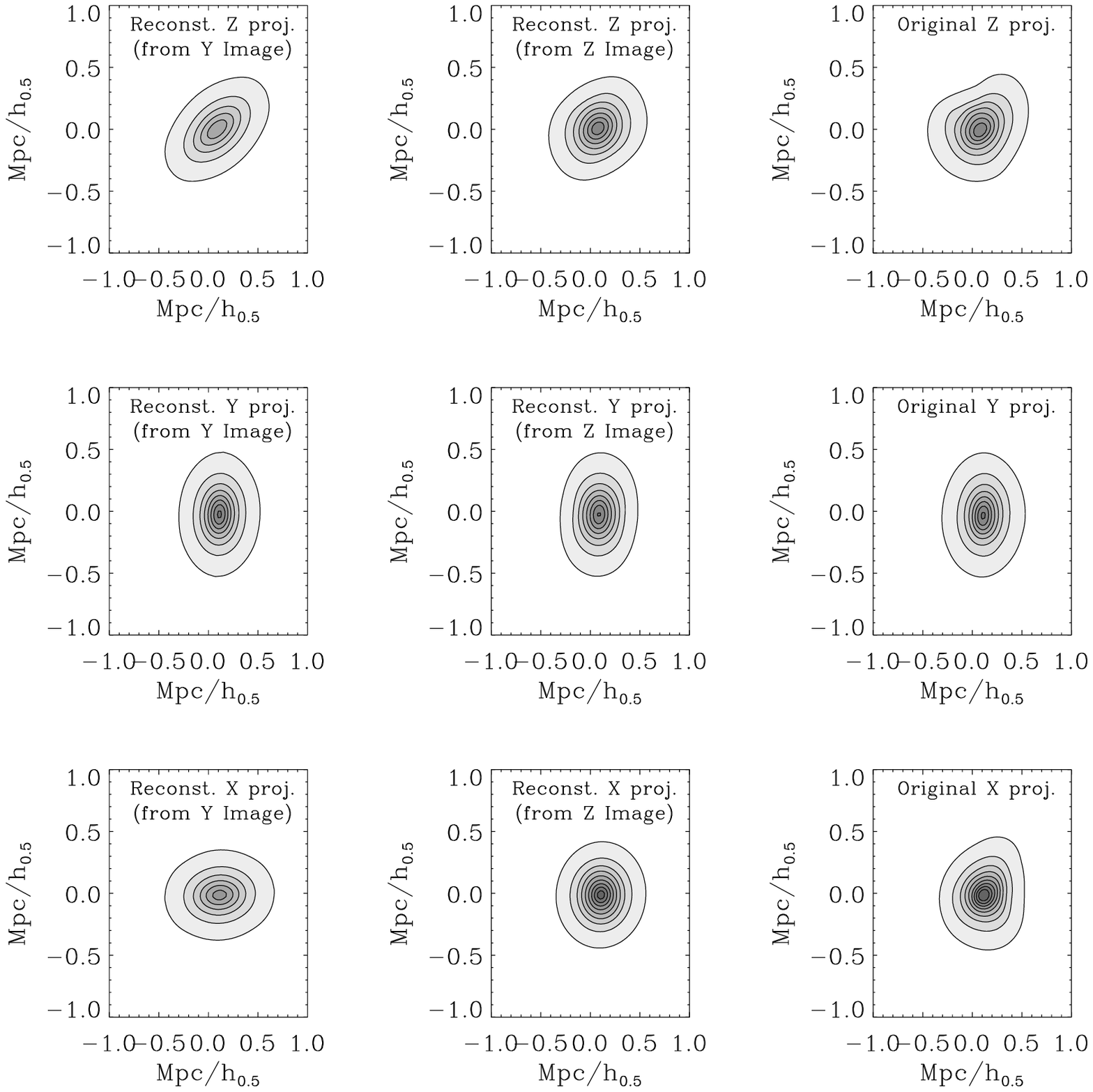}
\caption{The same as in Fig.~\ref{fig:projection_lens} but for the SZ maps}
\label{fig:projection_sz}
\end{figure*}

\begin{figure*}
\plotone{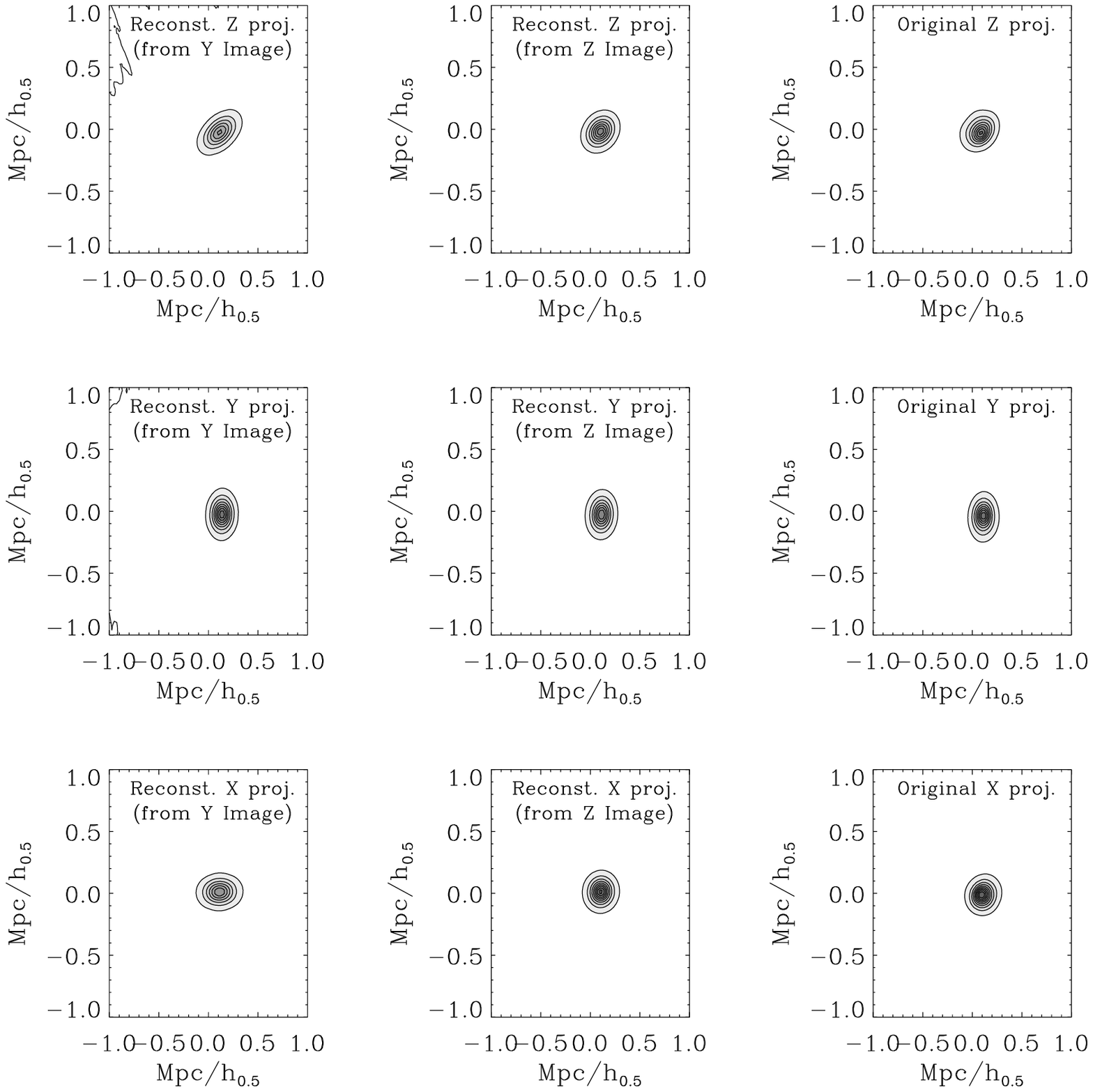}
\caption{The same as in Fig.~\ref{fig:projection_lens} but for the X-ray maps}
\label{fig:projection_xray}
\end{figure*}

Since we have three projections of the same underlying cluster, we can
make a more sophisticated comparison. We deproject the y-projections and
{\em predict} the two other orthogonal projections, searching for the
inclination angle that provides the best fit when compared with the true
x- and z-axis projections for the SZ, X-ray and surface density images.
We again find an inclination angle of $\theta_i=32^\circ$ for the
prolate case and $65^\circ$ for the oblate case.  A comparison between
the projected profiles and the real projections are shown in the left
panels of Figures \ref{fig:projection_lens}, \ref{fig:projection_sz} and
\ref{fig:projection_xray}. It is evident from those figures that the
reconstruction works very well apart from the innermost region where the
amplitude at the very center ($r\sim 0.2 \, \Mpc$) is underestimated by
$\approx 10\%$. Furthermore, due to heavy reliance on the elliptical
isothermal gas density distribution model for filling the COI, the
recovered 3D cluster distribution is somewhat over- elongated.
 
For an inclination angle of $32\deg$ the COI covers most of k-space and
one is forced to heavily rely on the assumed model to compensate for the
lost information. As another example we deproject the z-projection image
which has, for a prolate cluster, a rather large inclination angle,
$\theta_i\approx 84\deg$, and therefore has a negligible COI. The
reconstructions from the z-projection maps are shown in the left panels
of Figures \ref{fig:projection_lens}, \ref{fig:projection_sz} and
\ref{fig:projection_xray}. Due to the minor COI extrapolation in this
case the cluster center and geometry are better recovered and the other
two projections are well predicted, even in the innermost region of the
cluster.

\begin{figure*}
\plotone{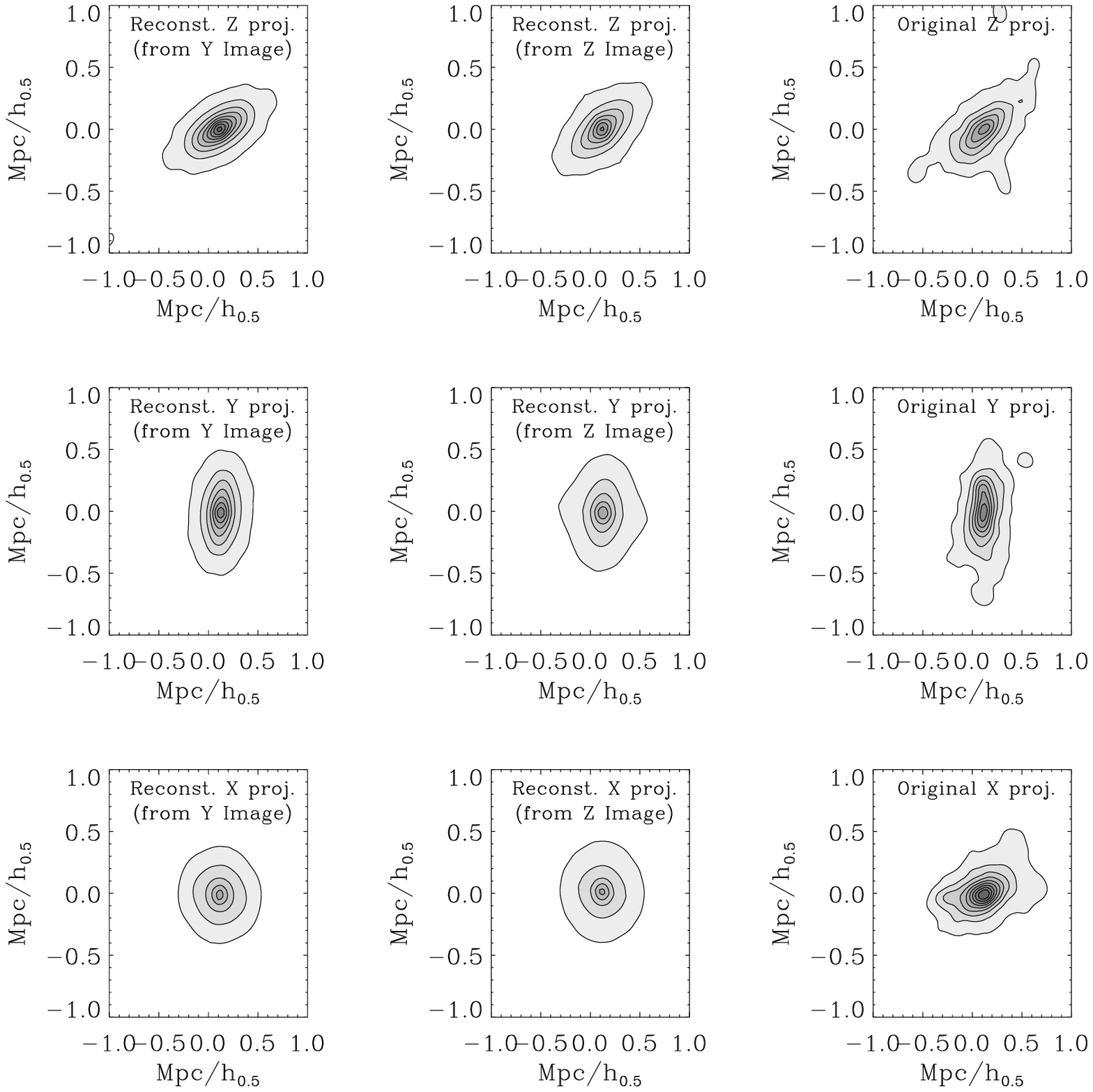}
\caption{The same as Fig.~\ref{fig:projection_lens}; but assuming an
oblate cluster}
\label{fig:obl_projection}
\end{figure*}

Figure~\ref{fig:obl_projection} shows the same comparison shown in the
left panels of Figure~\ref{fig:projection_lens} but for an oblate
cluster. In this case the reconstructed x-projection is rather poorly
predicted, supporting our contention regarding the cluster prolateness.

\subsubsection{Comparison with the True 3D Distributions}

\begin{figure*}[htbn]
\plotthree{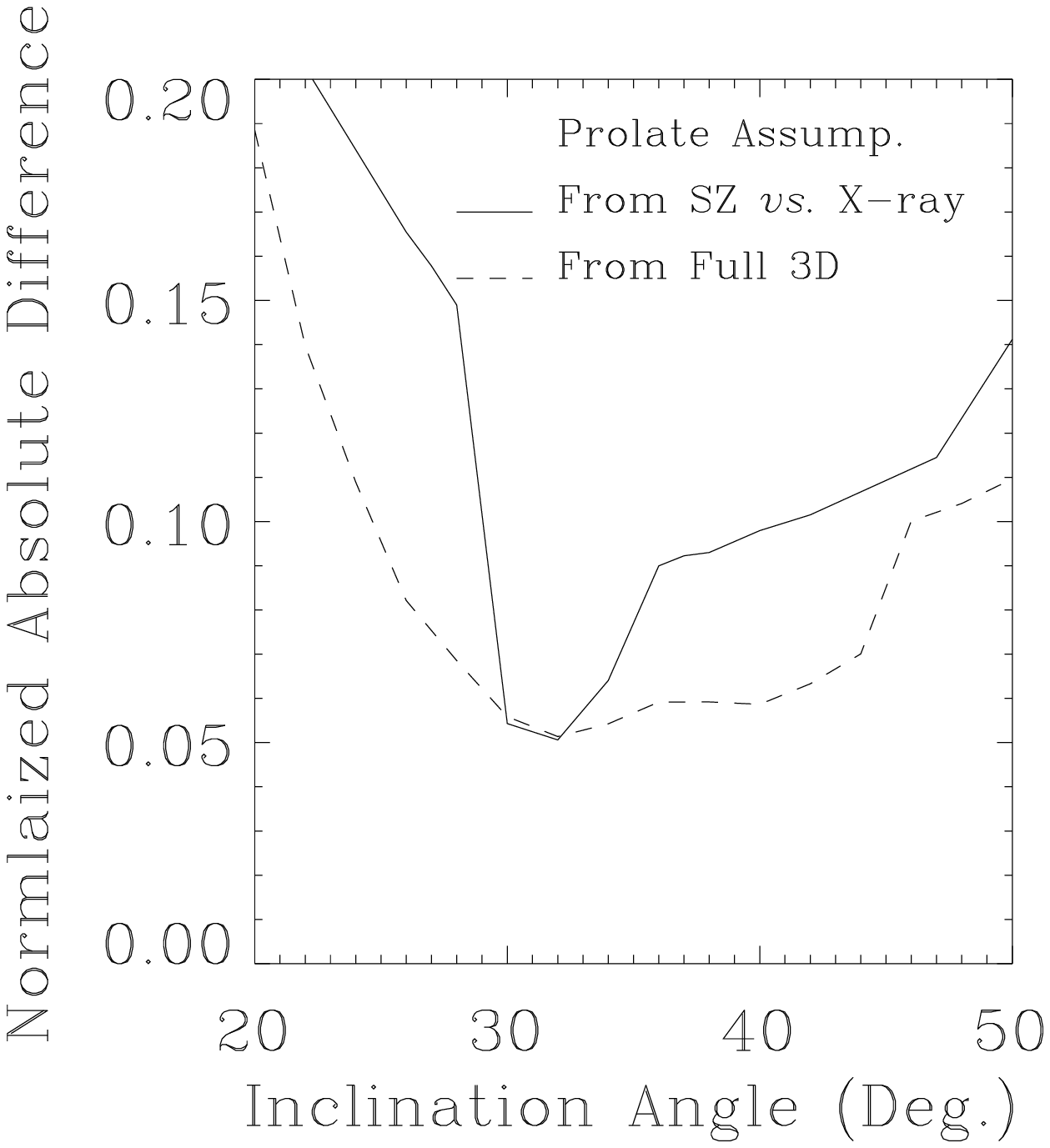}{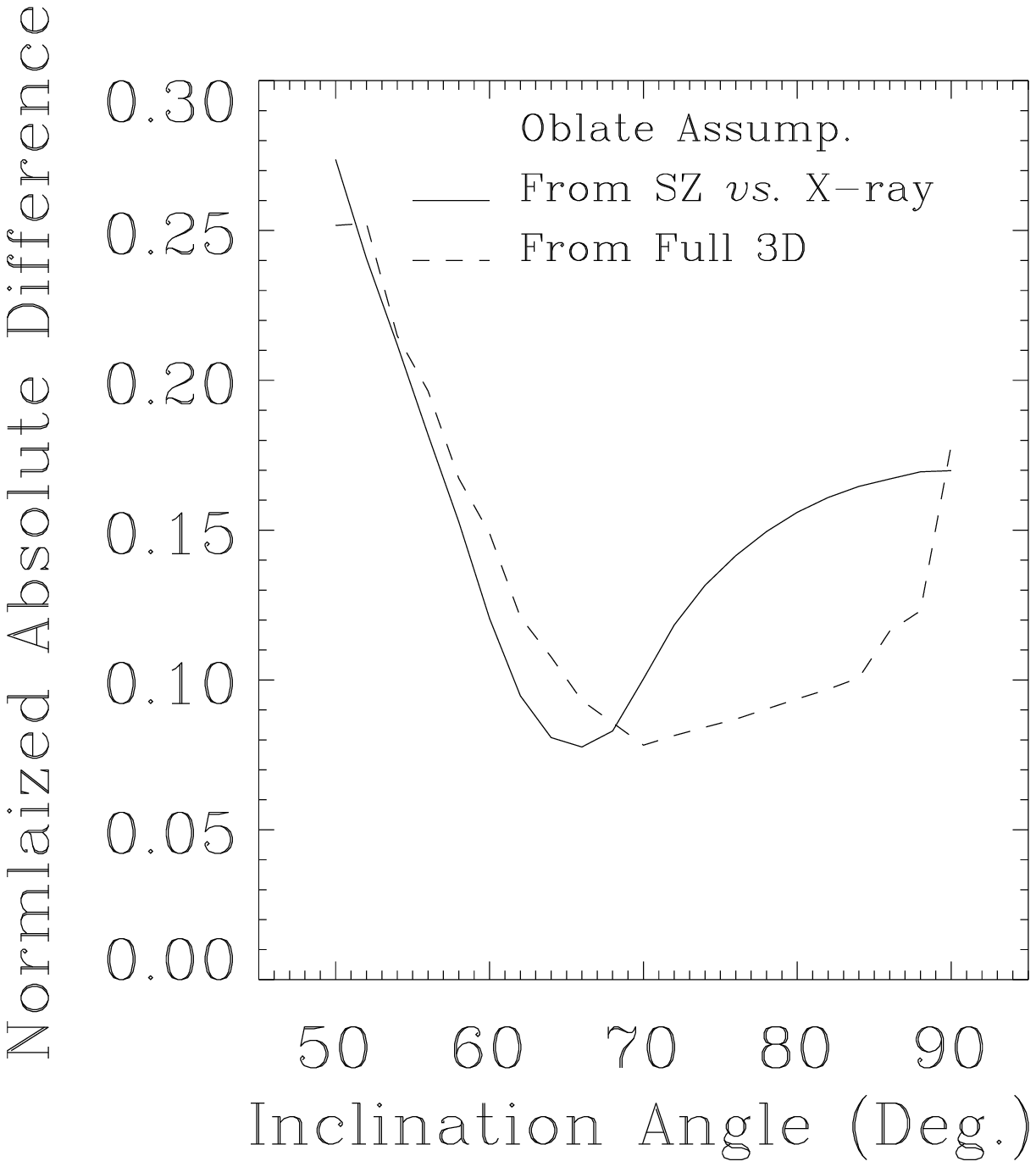}{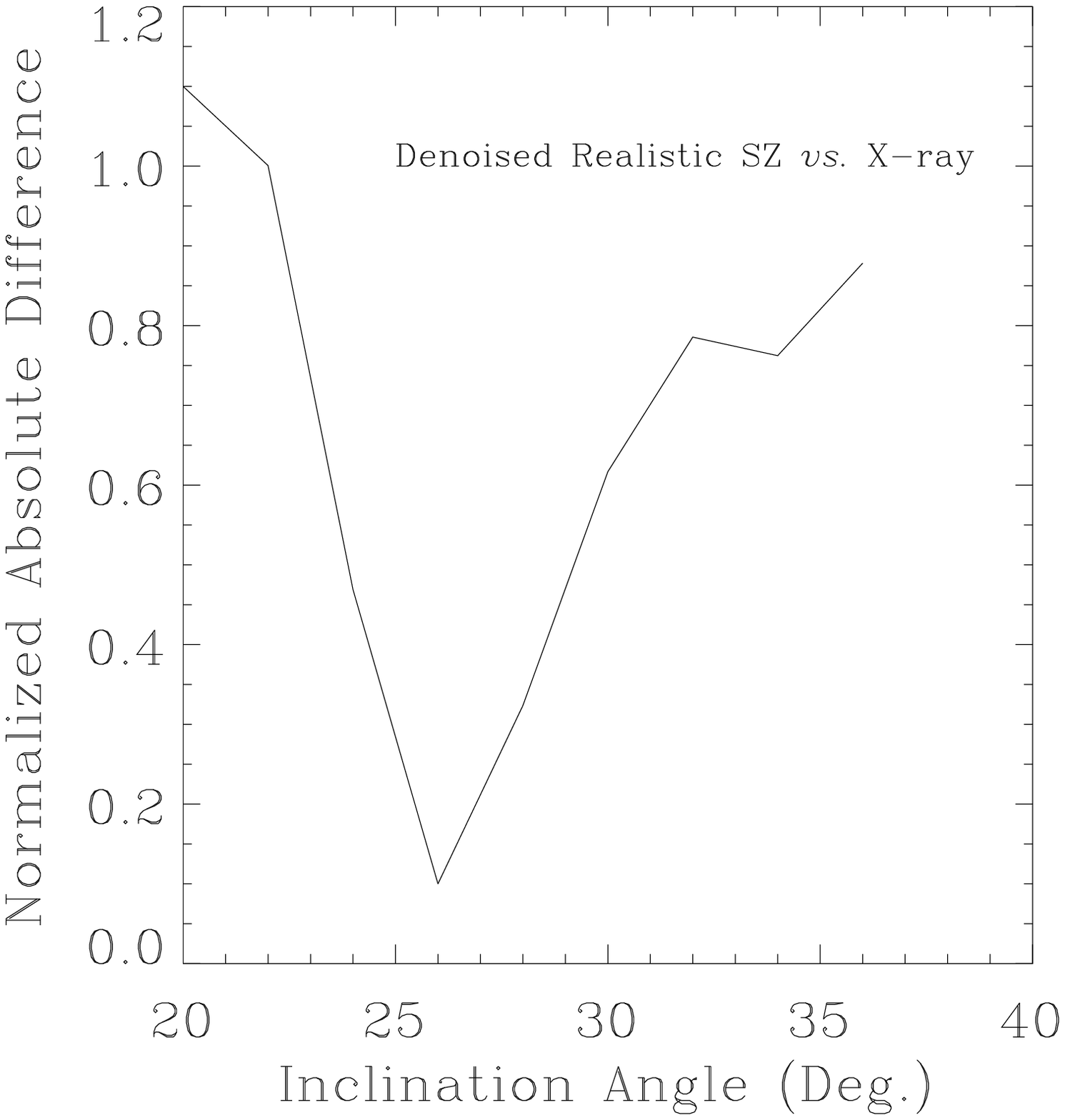}
\caption{The accuracy of the method: The dashed line displays the
cumulated, normalized absolute difference between the inferred and true
gas densities (where `true' is the gas density obtained directly from
our knowledge of the 3D distribution in the simulation data) as a
function of inclination angle. For comparison, we also display (solid
line) the difference between the gas density inferred from the X-ray
deprojection versus that from the SZ -- this mimics the procedure one
would do in practice with real data. Left and middle panels show the
result for prolate and oblate cluster assumption, respectively.  Right
panel: The same as before but with the comparison carried out with X-ray
and SZ images that have realistic noise and resolution levels (\S
\ref{sec:withnoise}).}
\label{fig:chi2vstheta}
\end{figure*}

Since we have the full 3D information from the simulations, we can
robustly test the accuracy of the deprojection by comparing the
deprojection with the true 3D gas or total mass distributions.  In
Figure \ref{fig:chi2vstheta}, we deproject the X-ray and SZ images as a
function of inclination angle. We transform the deprojected X-ray image
to calculate the gas density from the deprojection, and compare with the
true gas density by forming the normalized absolute difference:
\begin{equation}
\Delta(\theta_i) = \sum \frac{|\rho^{\rm X-ray}_{\rm gas}(r,z) - \rho^{\rm
true}_{\rm gas}(r,z)|} { |\rho^{\rm X-ray}_{\rm gas}(r,z) + \rho^{\rm
true}_{\rm gas}(r,z)| } ,
\end{equation}
where the $\rho^{\rm true}_{\rm gas}(r,z)$ is the true
(axi-symmetrized) density field in the simulation.
 
In the left and middle panels of Figure~\ref{fig:chi2vstheta}, the
dashed line displays the normalized absolute difference between the
inferred and true gas densities as a function of inclination
angle. For comparison, we also display (solid line) the difference
between the gas density inferred from the X-ray deprojection versus
that from the SZ -- this mimics the procedure one would do in
practice with real data, as discussed in \S
\ref{sec:determining_theta}. For this cluster, the agreement between
the two is superb, especially under the assumption of a prolate
cluster, demonstrating the reliability of the inclination angle
determination (within $\lsim \pm 5\deg$) even in the case of a very
wide COI.

\begin{figure*}[htbn]
\plottwo{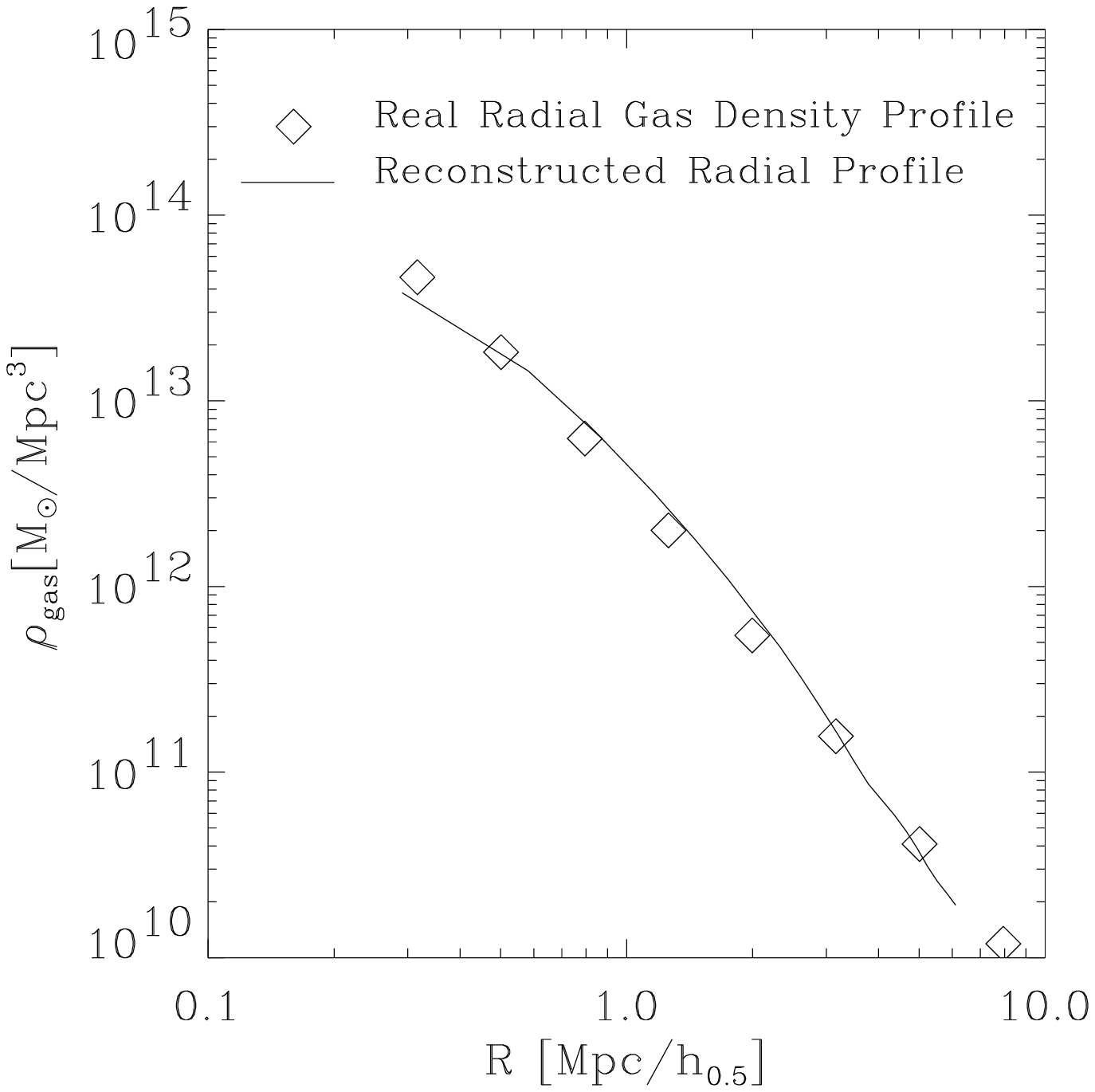}{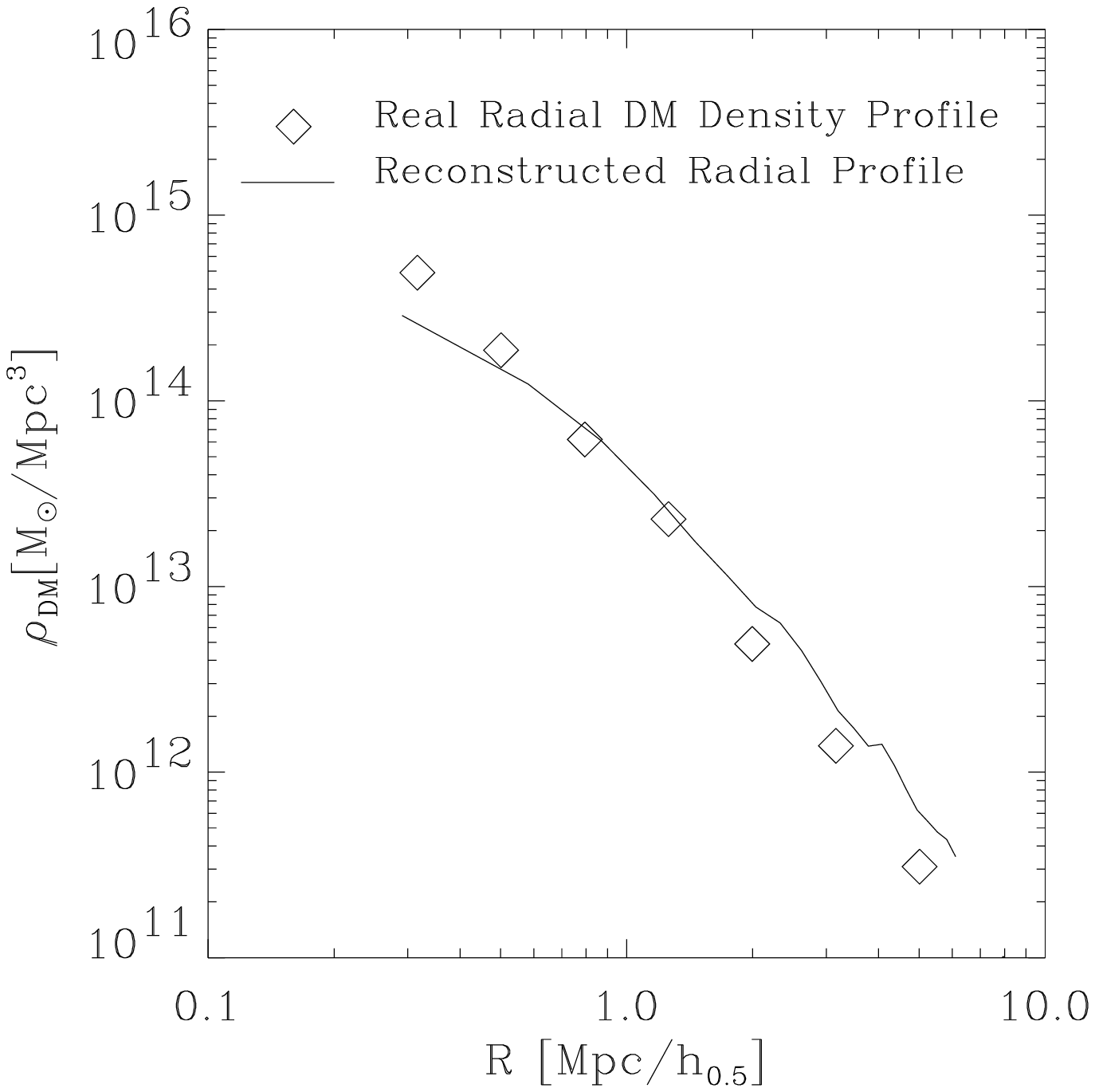}
\caption{ Left panel: The spherically averaged gas-density as a function
of radius as calculated from the 3D deprojected gas distribution (solid
line) and from the real (N-body) 3D distribution (diamonds). Right
panel: The same as in the left panel but for the dark matter radial
profile.}
\label{fig:spherical}
\end{figure*}

A useful way to quantify the quality of the deprojection is to is see
how well the radial profiles as calculated from the full and
deprojected 3D distributions compare. In Figure~\ref{fig:spherical} we
calculate the radial profiles of the gas distribution and dark matter
distribution.  In both cases the reconstruction underestimated the
density at the inner most center of the cluster but it yields a very
accurate profile at radii $\gtrsim 0.3 \; \Mpc$ for the gas density
profile and $\gtrsim 0.5 \, \Mpc$ for the dark matter density
profile. The discrepancy between the real and reconstruction radial
profiles in the innermost region, especially in the dark matter radial
profile, is primarily due to the loss of the high frequency information
in the COI, coupled with the finite resolution of the `observed' maps.
However, the {\em total} gas and mass differences are only of the order
of few percent.

\subsubsection{The Effect of Filling the COI on the Deprojection}
\label{rec:COI_effects}

In the full deprojection results shown until now, we have filled the COI
by extrapolating an isothermal ellipsoidal profile that best fits the
data outside the COI to within the COI. Here, we explore the effect of
two different schemes for filling the COI.  In
the first scheme we simply let the values within the COI to smoothly
(\ie\ exponentially) drop to zero. 
The second scheme is to simply perform
a linear extrapolation into the COI with the amplitude fixed by the
value at the cone boundary (see Figure~\ref{fig:coi}).

The general effect of first scheme is twofold. First, it results in a
decrease in the amplitude of reconstructed source function, with the
magnitude of the drop depending on the size of the COI. Second, it
produces a double lobed shape distribution.  As well, the second scheme
results in a decrease in the amplitude of the reconstructed source
function, and in more boxy shaped structures.

\begin{figure*}[htbn]
\plottwovert{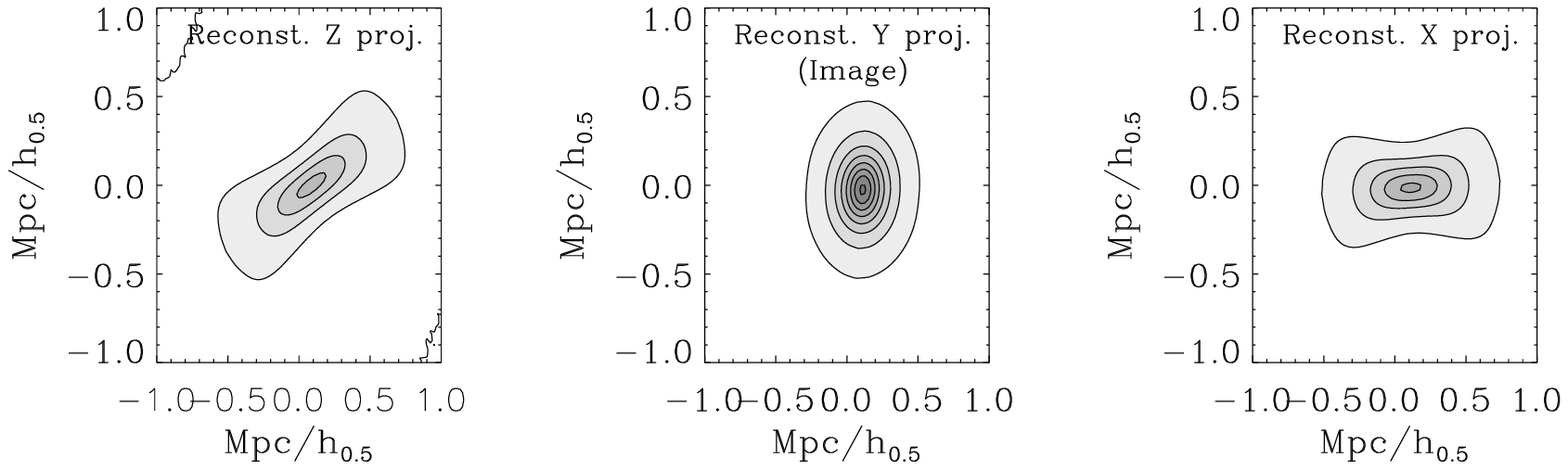}{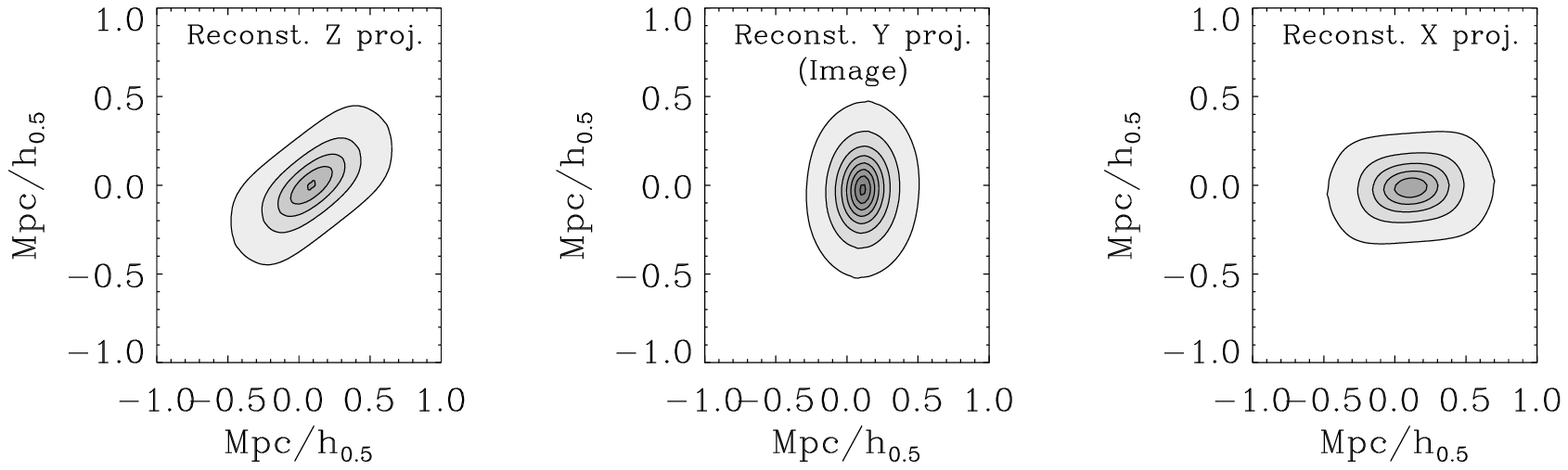}
\caption{ Upper row: Similar to the first column in
Figure~\ref{fig:projection_sz} but with the COI filled with values
fixed to its boundary. Lower row: The same as the upper panels but
with the COI filled with values that smoothly drop to zero.}
\label{fig:coi_test}
\end{figure*}

Figure~\ref{fig:coi_test} shows the {\it predicted} images obtained by
projecting the 3D reconstructed SZ source function into x-, y-, and
z-directions. Here we use the y-projected SZ as the input image for
the algorithm. The inclination angle here is $32^\circ$ which means
that the COI is quite wide. As expected, the re-projected y-image has
the same quality as the comparable image (first column and second row)
in Figure~\ref{fig:projection_sz}. The other two re-projected images
show in both cases a drop in the amplitude and in the case of
``constant filling'' the reconstruction has a more boxy features while
the ``zero filling'' results in a double lobed reconstruction.

\subsubsection{Addition of Realistic Noise and Spatial Resolution}
\label{sec:withnoise}

We have demonstrated thus far that our procedure for filling the COI and
comparing the X-ray and SZ maps is sufficient to uniquely determine the
inclination angle and hence determine the underlying 3D density
structure. However, our calculations thus far have employed essentially
infinite spatial resolution, and have been noise-free. In order to test
if the method will be viable with realistic noise and instrumental
response, we performed the following simple simulation: The SZ maps have
been degraded to obtain $50^{\prime\prime}$ spatial resolution and peak
$S/N \simeq 20$ similar to the resolution and sensitivity of the SZ
observation attained at the OVRO and BIMA arrays
(\cite{carlstrom95}). Similarly, the X-ray images have been degraded to
spatial resolution of $10^{\prime\prime}$, and peak $S/N \simeq 10$,
mimicking the resolution and sensitivity of CHANDRA for a rich cluster
of galaxies. 

The right panel of Figure \ref{fig:chi2vstheta} shows the angle
determined from the X-ray {\it vs.} SZ maps comparison. The maps are
initially denoised with a wavelet denoising algorithm (see Paper I;
Brosch \& Hoffman 1999; Hoffman 2000) and then the deprojected X-ray and
SZ images are compared as a function of inclination angle. The best fit
inclination angle in this case is $26^\circ$ which is within $6^\circ$
from the angle obtained from noiseless images; this is an excellent
agreement that shows the potential of this method when applied to real
data.

\section{Testing the Method with the Full Simulated Cluster Sample}

\begin{figure*}
\plotone{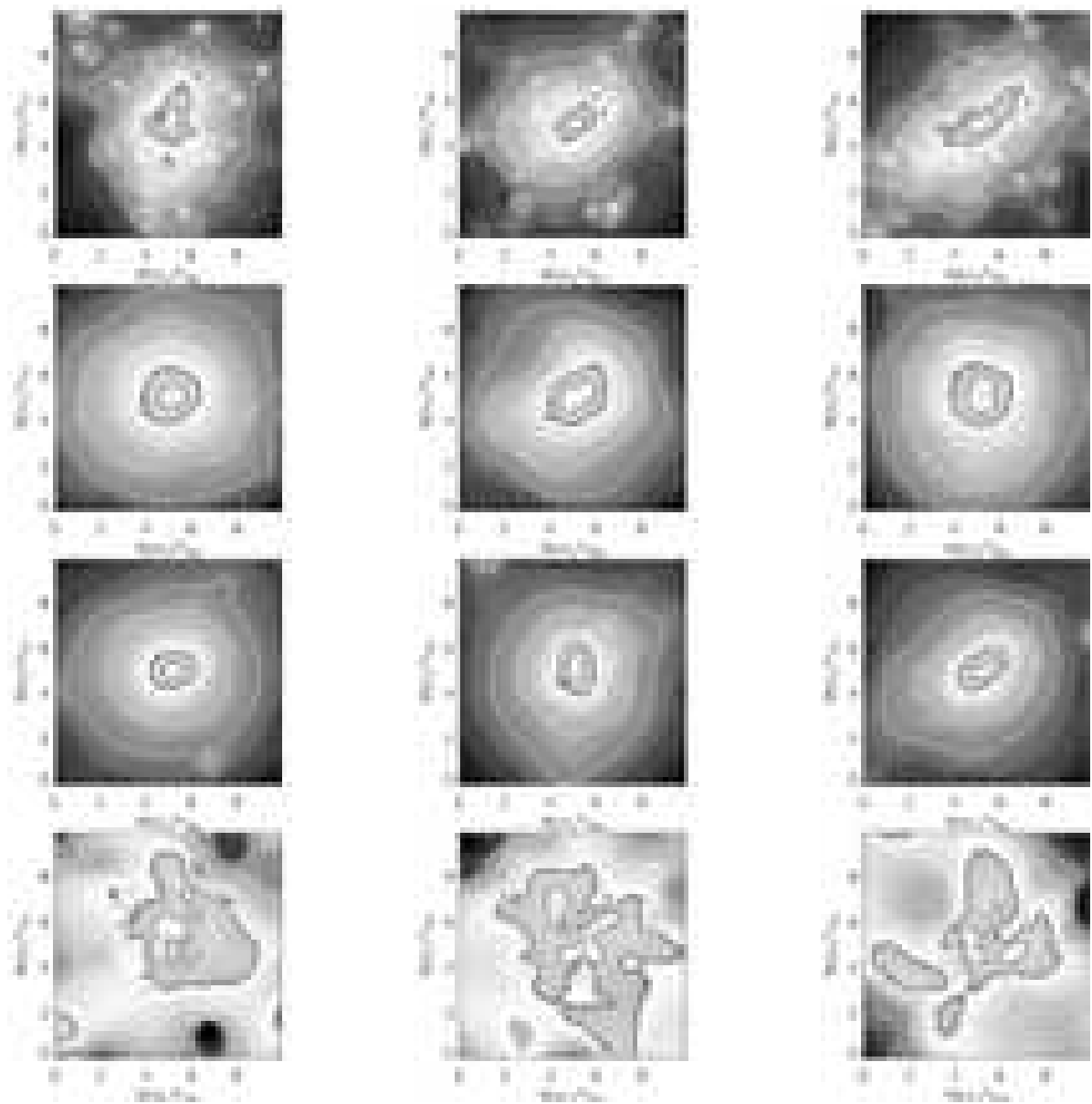}
\caption{Left column: The x-projection for the prototype cluster at
$z=0.3$. The grey scale and contours show the logarithmic
intensity. Surface density (first panel; $M_\odot/{\rm Mpc}^2$), and
Sunyaev-Zel'dovich decrement (second panel). X-ray surface brightness
(third panel; $M_\odot^2 K^{1/2} /{\rm Mpc}^5$), and emission weighted
temperature (fourth panel; $K$). The middle and right columns display
the same plots, but for the y- and z-projection maps respectively.}
\label{fig:allproj_z03}
\end{figure*}

\begin{figure*}
\plotone{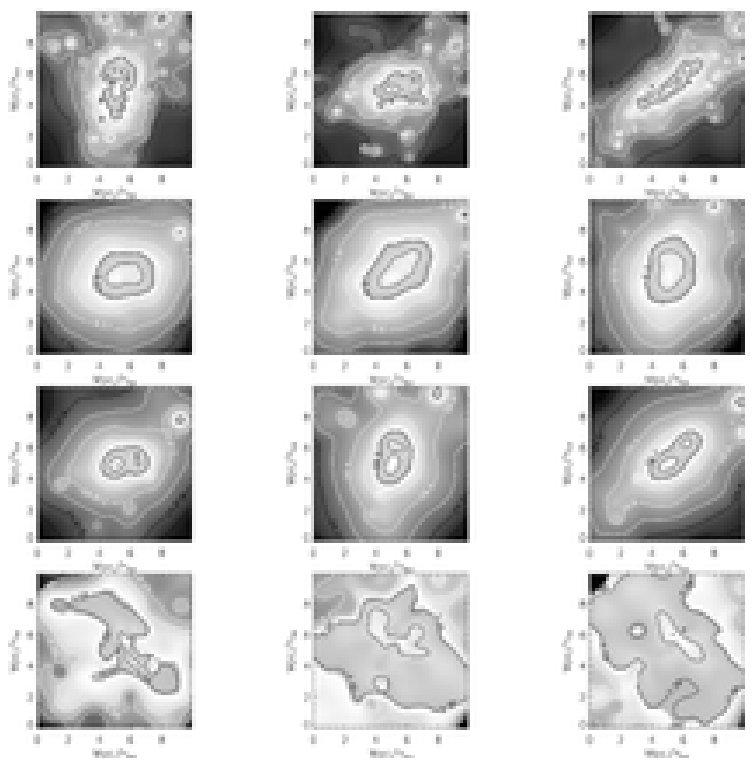}
\caption{Left column: The x-projection for the prototype cluster at
$z=0.6$. The grey scale and contours show the logarithmic
intensity. Surface density (first panel; $M_\odot/{\rm Mpc}^2$), and
Sunyaev-Zel'dovich decrement (second panel). X-ray surface brightness
(third panel; $M_\odot^2 K^{1/2} /{\rm Mpc}^5$), and emission weighted
temperature (fourth panel; $K$). The middle and right columns display
the same plots, but for the y- and z-projection maps respectively.}
\label{fig:allproj_z06}
\end{figure*}

\begin{figure*}
\plotone{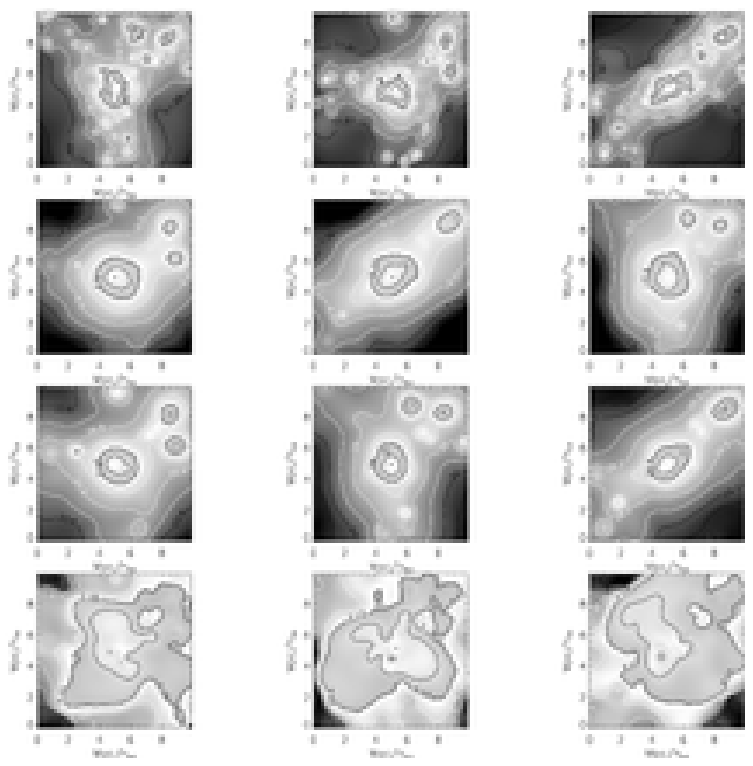}
\caption{Left column: The x-projection for the prototype cluster at
$z=0.9$. The grey scale and contours show the logarithmic
intensity. Surface density (first panel; $M_\odot/{\rm Mpc}^2$), and
Sunyaev-Zel'dovich decrement (second panel). X-ray surface brightness
(third panel; $M_\odot^2 K^{1/2} /{\rm Mpc}^5$), and emission weighted
temperature (fourth panel; $K$). The middle and right columns display
the same plots, but for the y- and z-projection maps respectively.}
\label{fig:allproj_z09}
\end{figure*}

In Figure \ref{fig:allproj_z03} we display the logarithmically scaled
mass surface density, X-ray surface brightness, SZ decrement and
emission weighted temperature distributions for the three orthogonal
projections of the cluster at $z = 0.3$. As before, the dimensions for
the surface density, temperature and SZ decrement are physical
($M_\odot/{\rm Mpc}^2$ and $K$ for the first two, while the SZ decrement
is dimensionless), while we recall that the X-ray surface brightness was
calculated as the line of sight integral $\int \rho_{\rm gas}^2 T^{1/2}
dl$ and has dimensions $M_\odot^2 K^{1/2} /{\rm Mpc}^5$. Figures
\ref{fig:allproj_z06} and \ref{fig:allproj_z09} show the same plots for
the cluster at $z = 0.6$ and $z = 0.9$ respectively.

The merger history of the cluster can be traced through this series of
simulation outputs. At $z=0.9$, the central mass concentration is in
place, while two large sublumps are infalling at a projected co-moving
distance of $\simeq 4 \, \Mpc$ from the cluster center.  At $z = 0.6$,
the entire system undergoes a major merger event. The three main mass
concentrations are most clearly seen in the cores of the x- and
y-projections (Figures \ref{fig:allproj_z09} and
\ref{fig:allproj_z06}), while the morphology of both the gas and dark
matter in the z-projection becomes highly elliptical (Figure
\ref{fig:allproj_z03}). This particular simulation output provides the
most stringent test for the deprojection method: the system is clearly
far away from dynamical equilibrium and certainly is very poorly
described by a spherical model for the underlying gas and dark matter
distributions. Nevertheless, as we shall demonstrate, the system is
fairly well described as axially symmetric; the symmetry is qualitatively
evident in the SZ and X-ray images, and, to a lesser extent, in the
surface mass distributions.

\begin{figure*}
\plotone{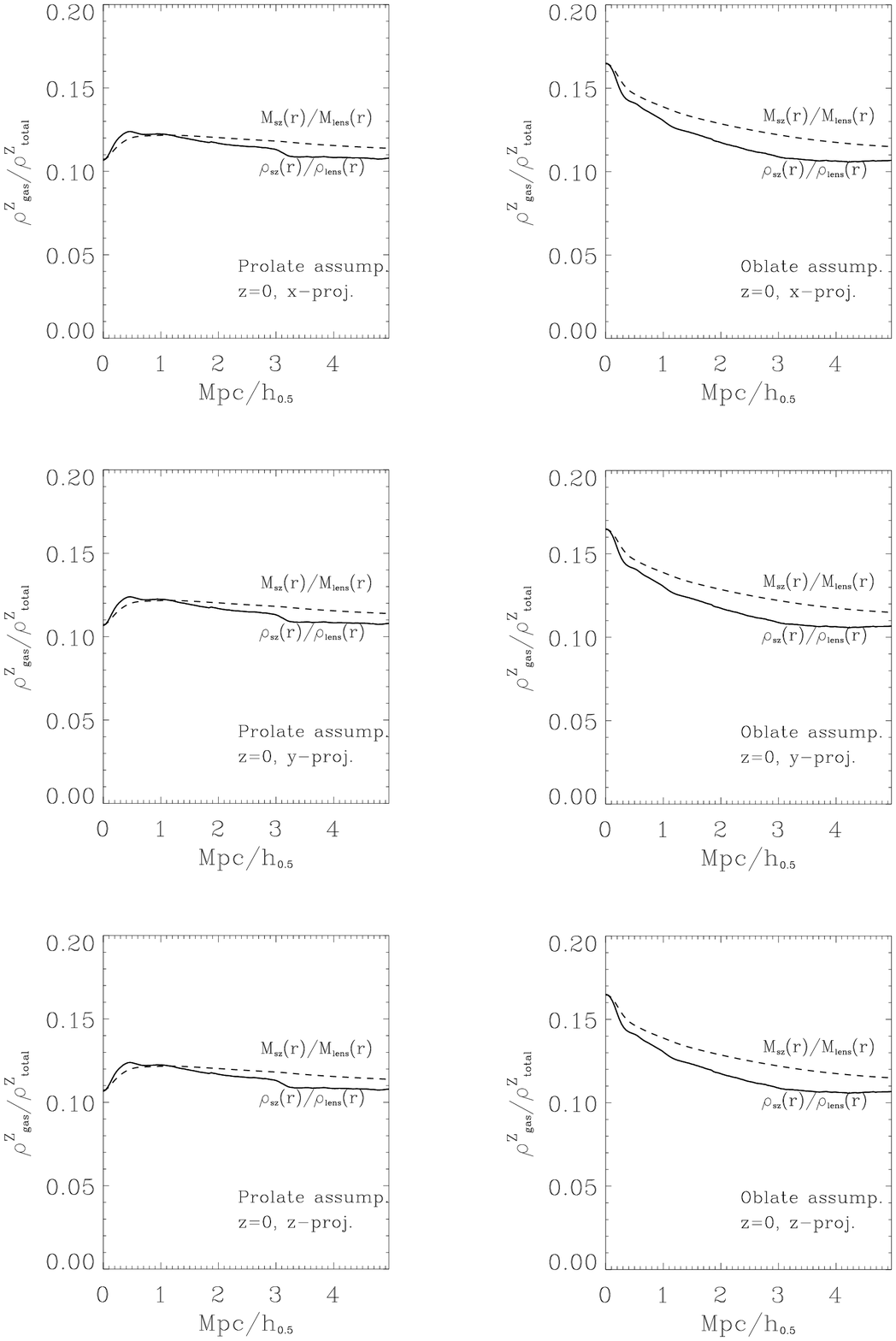}
\caption{The ratio of the z-integrated gas density and dark matter
density, $\rho^z_{gas}/\rho^z_{DM}$ (solid lines), and the corresponding
gas to dark matter mass ratio (dashed line), as a function of r,
calculated from the surface density and Sunyaev-Zel'dovich maps of the z
= 0 output.  The left and right panels are produced assuming a prolate
and oblate cluster respectively.}
\label{fig:z00_densityratio}
\end{figure*}

\begin{figure*}
  \plotone{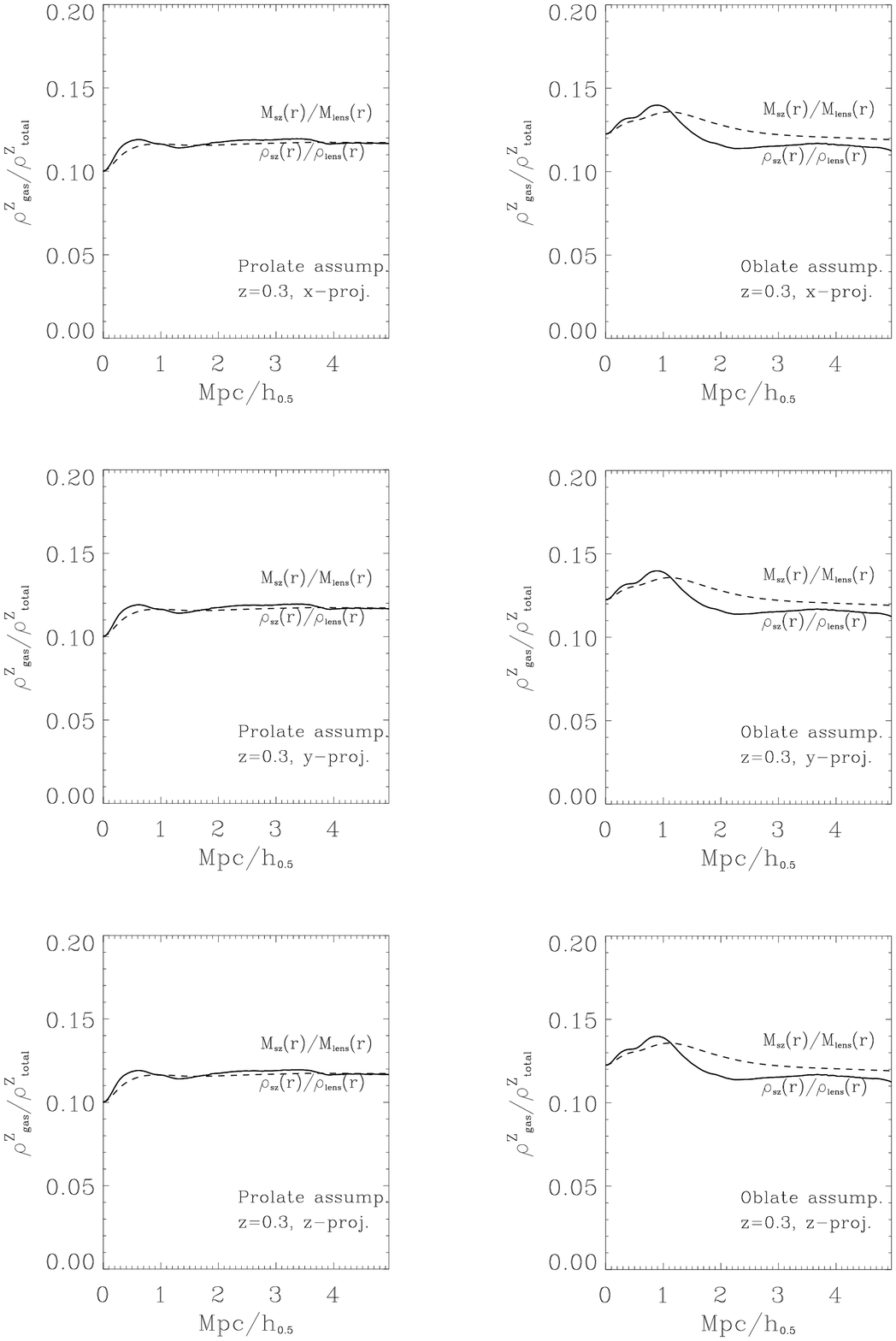}
\caption{As in Figure \ref{fig:z00_densityratio} but for the z = 0.3
 output.}
\label{fig:z03_densityratio}
\end{figure*}

\subsection{The Baryon Fraction for All Clusters} 

\begin{figure*}
\plotone{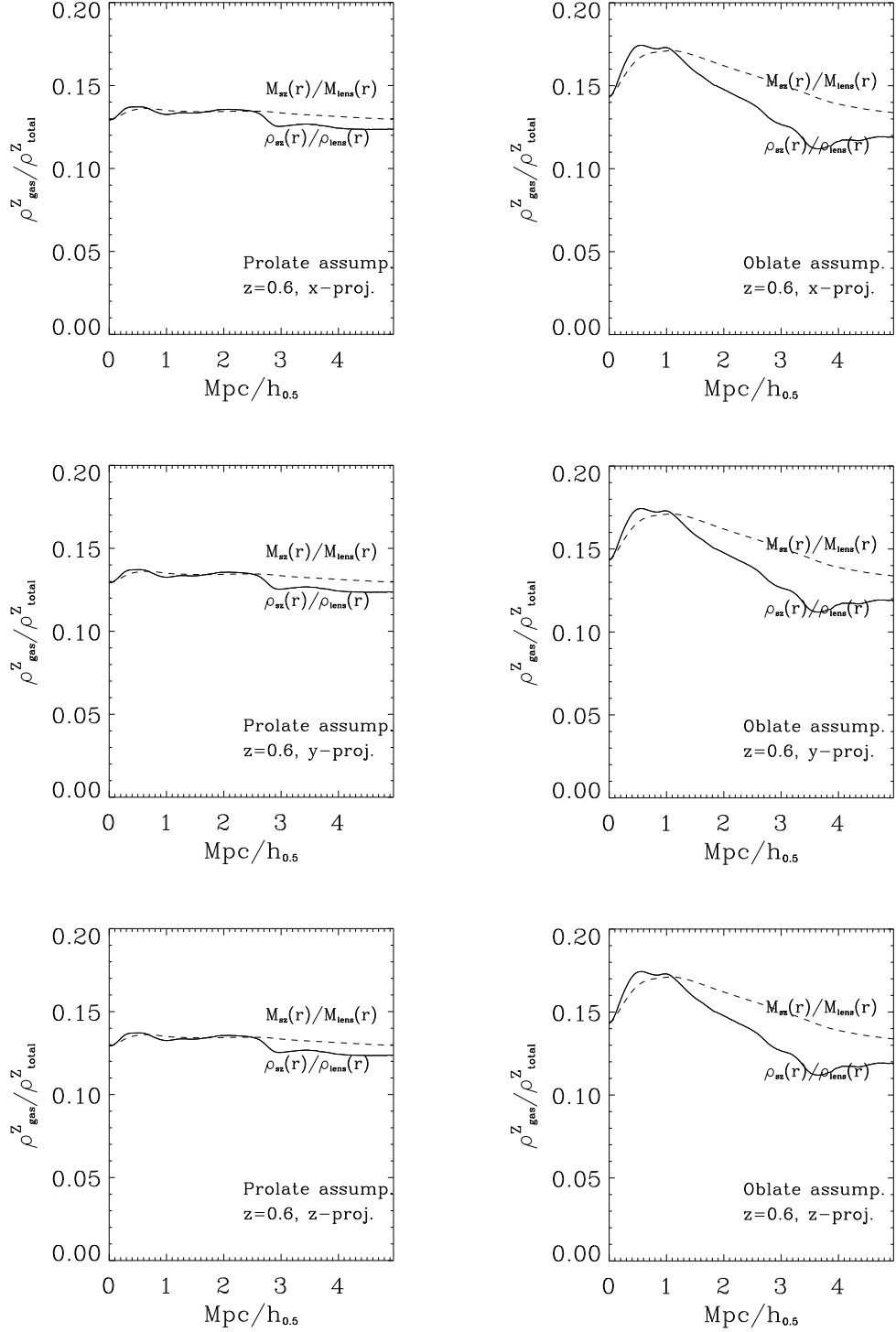}
\caption{As in Figure \ref{fig:z00_densityratio} but for the z = 0.6
 output.}
\label{fig:z06_densityratio}
\end{figure*}

\begin{figure*}
\centering
\plotone{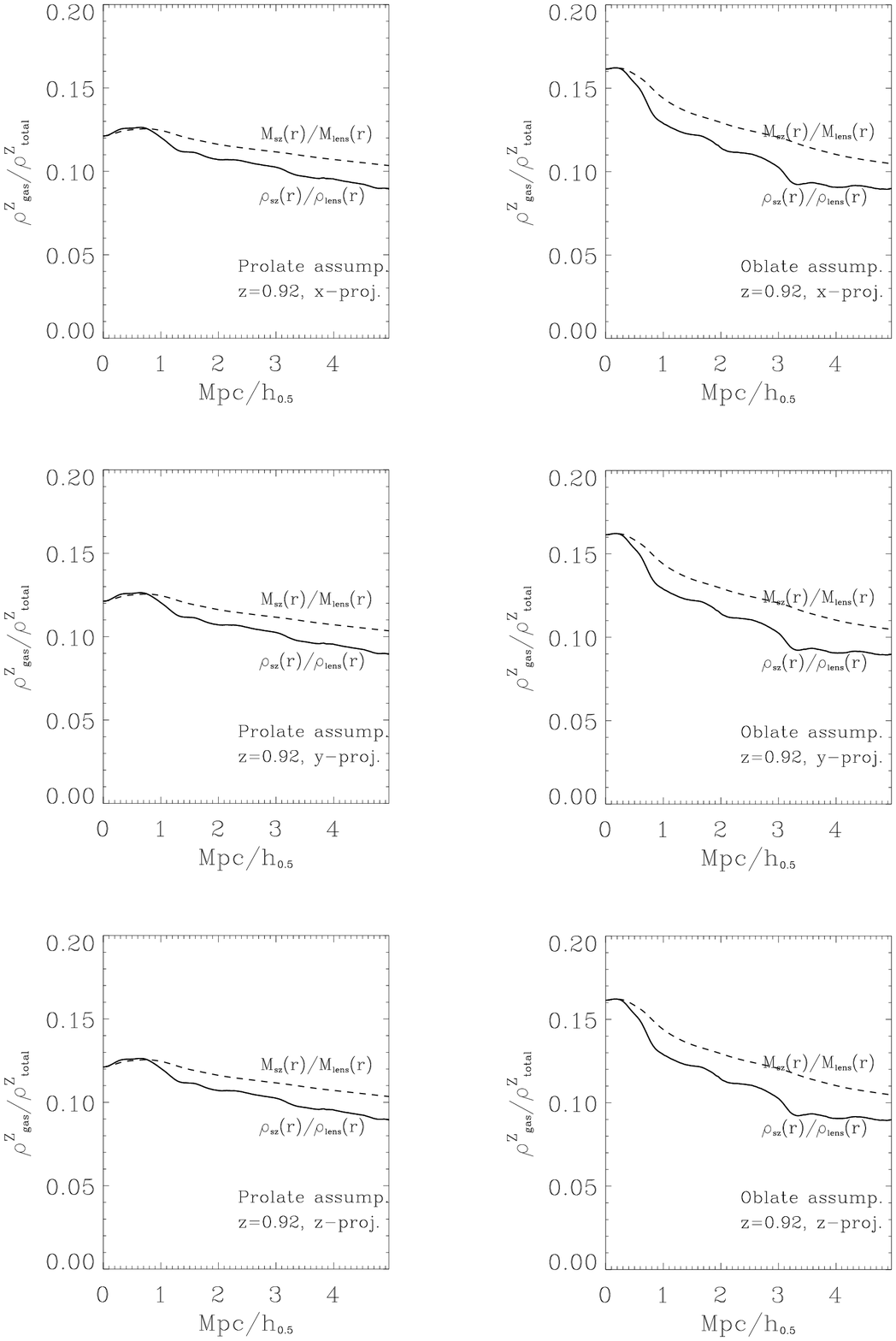}
\caption{As in Figure \ref{fig:z00_densityratio} but for the z = 0.9
output.}
\label{fig:z09_densityratio}
\end{figure*}

Using equation (\ref{eqn:rhoz}), we calculated the gas to total mass and
density ratios for the full suite of cluster outputs. The results are
shown in Figures \ref{fig:z00_densityratio}, \ref{fig:z03_densityratio},
\ref{fig:z06_densityratio}, and \ref{fig:z09_densityratio} for the
simulation outputs at $z = 0.0, 0.3, 0.6, \;{\rm and} \; 0.9$
respectively.

In all cases, the method recovers very well the simulation input baryon
fraction. The radial dependence of the gas to total mass ratio is
flatter under the assumption of prolateness in the underlying 3D
structure.
Therefore, one can possibly use the expected small variability in gas
to mass ratio to determine whether the cluster is prolate or oblate.
In all the cases we examined, the gas to mass ratio shows less
variability when its shape is assumed to be prolate, which is in total
agreement with the real distribution of the gas and mass in those
clusters.

\subsection{The Inclination Angle} 

\begin{figure*}[htbn]
\centering \leavevmode
\epsfxsize=0.5\textwidth \epsffile{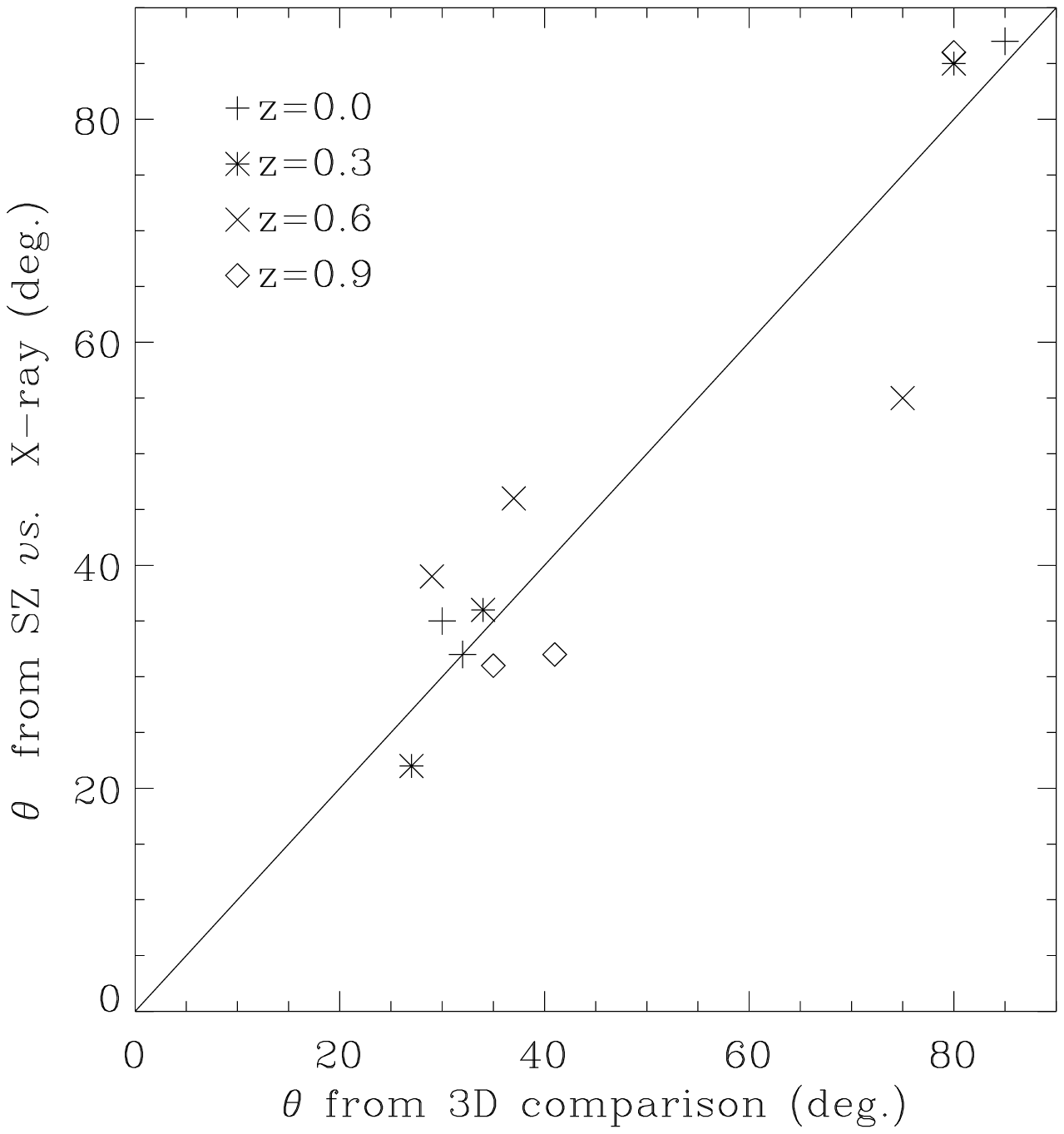}
\caption{The inclination angle as determined from the SZ {\it vs.}
X-ray for each of the x,y, and z-projection images at  four
redshifts (12 clusters) drawn against the inclination angle as
determined from the full real 3D distribution. The cluster at each
redshift is marked with different symbols.}
\label{fig:angle_det}
\end{figure*}

The inclination angle determination from the cluster SZ {\it vs.} X-ray
images at the high redshifts is somewhat more complicated. This
complication stems from the fact that at higher redshifts the cluster
is not very relaxed, especially at $z=0.6$ where the cluster undergoes
a major merger, rendering the connection between the images and the
temperature uncertain. Nevertheless, we have attempted to reconstruct
assuming a constant cluster temperature, chosen to be the mean
temperature in the region around the cluster center with `intensities'
larger the $30 \%$ of the image maximum density. We note that the
scatter around that mean temperature in the chosen region is about
$15\%$ at worst. Figure~\ref{fig:angle_det} shows the inclination
angle as determined from a comparison of the SZ and X-ray images as
viewed from the x, y and z-projections and at four different redshifts
{\it vs.} the inclination angle as determined from the actual 3D
distribution. The agreement between the two is very good.

\begin{figure*}
\plotone{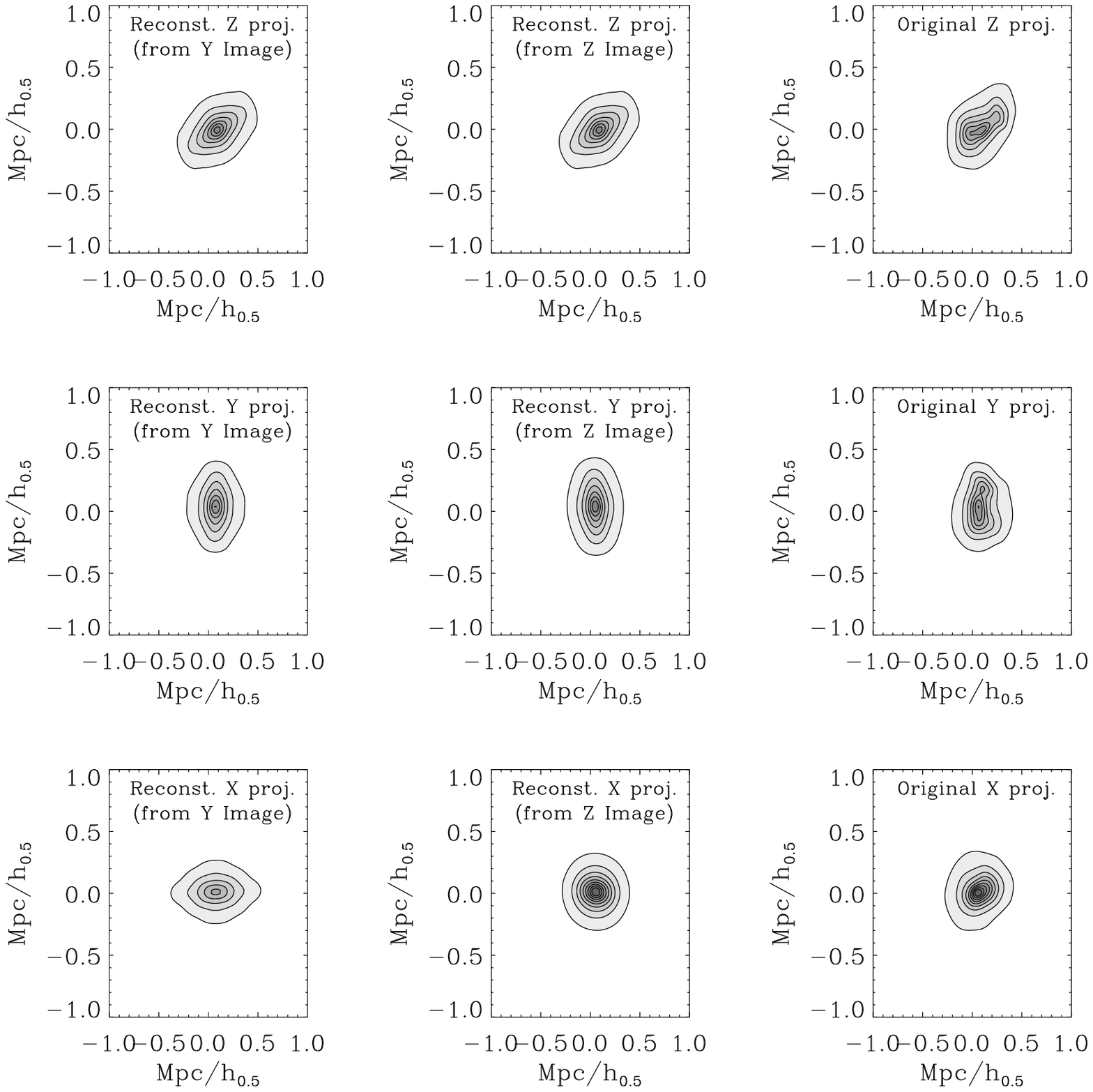}
\caption{Similar to Fig.~\ref{fig:projection_sz} (reconstruction from
y \& z-projection SZ maps); but with the cluster at redshift 0.3}
\label{fig:projection_sz3}
\end{figure*}

\begin{figure*}
   \plotone{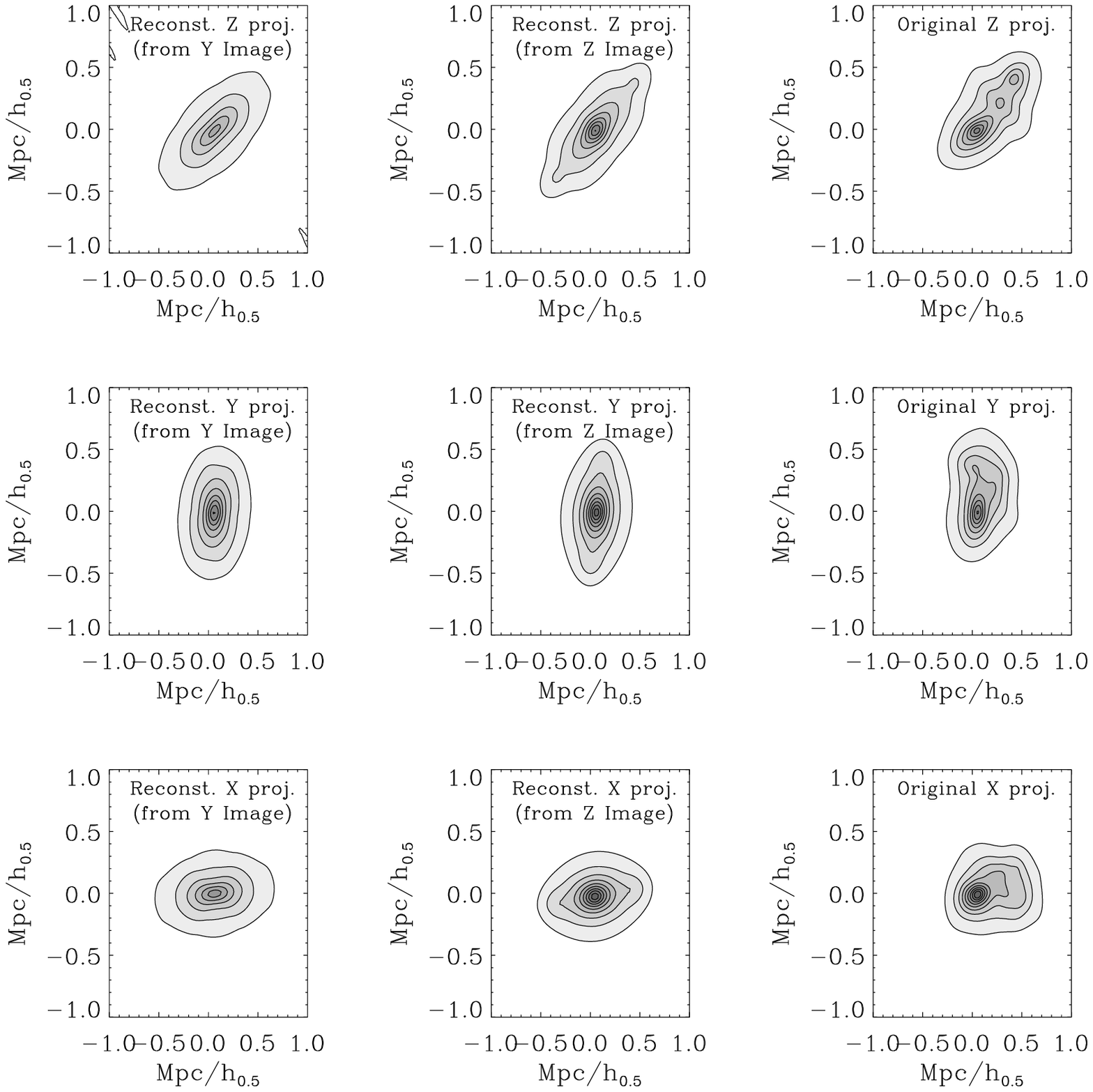}
  \caption{Similar to Fig.~\ref{fig:projection_sz} (reconstruction from
  y \& z-projection SZ maps); but with the cluster at redshift 0.6}
\label{fig:projection_sz6}
\end{figure*}

\begin{figure*}
  \plotone{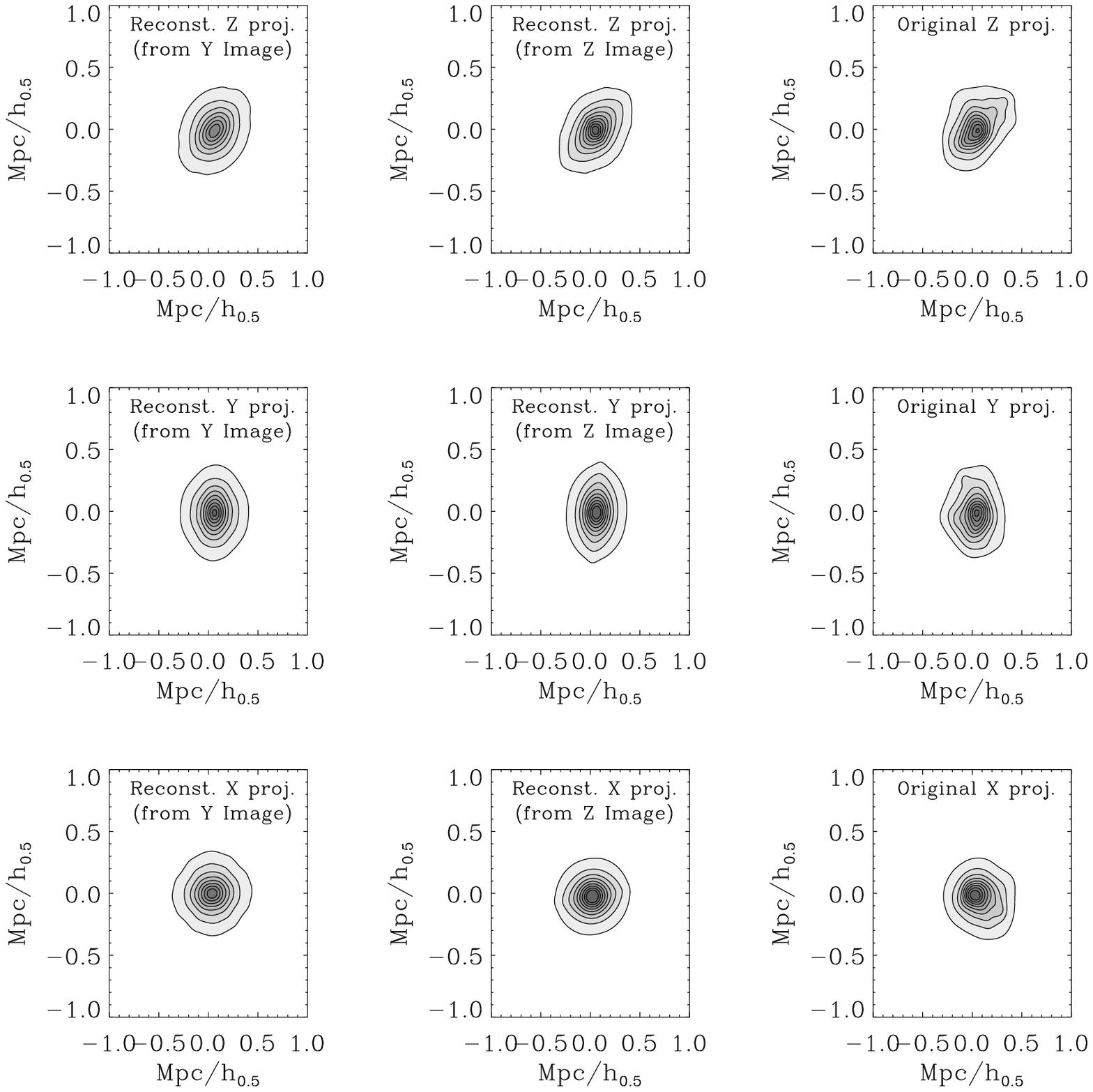}
  \caption{Similar to Fig.~\ref{fig:projection_sz} (reconstruction from
  y \& z-projection SZ maps); but with the cluster at redshift 0.9}
  \label{fig:projection_sz9}
\end{figure*}

Obviously, filling the COI with an elliptical isothermal density becomes
unrealistic for the $z=0.6$ case, where a major merger is taking
place. In principle, one can vary the density model adopted for filling
the COI. Here we choose to stay with the isothermal model with the hope
of obtaining a reasonable reconstruction the cluster central
region. Figures~\ref{fig:projection_sz3},~\ref{fig:projection_sz6} and
~\ref {fig:projection_sz9} shows how well the method performs in the
various redshifts when the z-projection images are used for
reconstruction.  Note the reconstruction is reasonable in all cases with
the exception of the $z=0.6$ case where the details are different.

\section{Discussion}

We have tested a method for deprojecting realistic galaxy cluster X-ray,
SZ decrement and mass surface density maps, assuming only that the
underlying gas/mass distributions are axially symmetric.

Conceptually, the deprojection procedure is straightforward: we first
determine the image axis of symmetry, employing three different
estimators accounting for the Poisson nature of the noise, and a
technique giving increased weight to counts at large radii from the
cluster center. Having a solution for the projected symmetry axis, we
deproject the SZ, X-ray and surface mass distributions using equation
(\ref{eqn:maineqn}), allowing for the possibility that the cluster is
either oblate or prolate.  As the inclination angle is the primary
unknown, we deproject each image over a wide range of assumed
inclination angles, searching for the angle that consistently yields the
best reconstructed ratios between the Sunyaev-Zel'dovich, X-ray and the
total mass deprojections.

At each inclination angle, we fill the information lost in the COI by
either an elliptical isothermal model fitted to the data outside the
COI, or a linear extrapolation into the COI with the amplitude fixed by
the value at the cone boundary.  The results are fairly insensitive to
the precise algorithm applied to fill the COI, primarily because the
information lost inside the COI is restricted to moderately high spatial
frequencies. 

For this study, we demonstrated this procedure in depth for a single,
prototype cluster, and then applied the method to the entire sample.  We
demonstrated the goodness and robustness of the inclination angle
determination in several ways: we compared the 3D gas density
distribution as inferred from the SZ and X-ray deprojection, using the
appropriate best fit angle for each case. The agreement between the two,
in terms of both shape and amplitude, is very good (with a $5-10\%$
difference at the center).

We also determined the inclination angle by calculating the cumulated,
normalized absolute difference between the inferred and true gas
densities as a function of inclination angle (see
Figure~\ref{fig:chi2vstheta}). The results agreed with our previous
determinations to an accuracy of $\lsim \pm 5\deg$, demonstrating the
reliability of the inclination angle determination even in the case of a
very wide COI.

With the prototype cluster, we mimic observations of a rich galaxy
cluster by CHANDRA (X-ray) and BIMA or OVRO (SZ) by degrading the
resolution of the simulated data and adding noise. We again compared
the deprojections of the two maps to determine the inclination angle
and found $26^\circ$ (assuming a prolate cluster) which is within
$6^\circ$ from the angle obtained from noiseless images; this is an
excellent agreement that shows the potential of this method when
applied to real data.  We note here, that the purpose of this exercise
is to show that the method has the potential of reproducing the 3D
structure from the current data sets, this however does not mean that
we have tested the application of the method to those data sets. The
application of the method to each of those data sets and its unique
specifications should be tested much more extensively, something we
intend to carry out in a forthcoming study.

   We also note, that since the current method is based on Fourier
slice theorem it is well suited to analyze interferometric data, like
the data obtained by the BIMA and OVRO experiments, where the raw data
is given in the Fourier space U-V plane. For this kind of data one can
for example, manipulate the X-ray or lensing data and transform them to
the Fourier U-V plane and make the comparison with the SZ raw data at
the wavenumbers covered in this plane which has the advantage of
avoiding the U-V plane extrapolation needed in order to obtain a real
space SZ image.

Finally, with the prototype cluster we tested the accuracy of the
deprojection by comparing the true and inferred 3D radial profiles of
the gas and total mass distributions (Figure~\ref{fig:spherical}). We
found that the reconstruction underestimated the density at the
innermost center of the cluster but it yields a very accurate profile
at radii $\gtrsim 0.3 \; \Mpc$ for the gas density profile and
$\gtrsim 0.5 \, \Mpc$ for the dark matter density profile. The
discrepancy between the real and reconstruction radial profiles in the
innermost region, especially in the dark matter radial profile, is
primarily due to the loss of the high frequency information in the
COI. However, the {\em total}, estimated over the whole the simulation
volume, gas and mass differences are only of the order of few percent.

We also developed a feature of the method that provides a direct
determination of the baryon fraction in clusters, independent of the
cluster inclination angle. The only uncertainty comes from whether the
underlying cluster gas density distribution is assumed to be prolate or
oblate. For all of the clusters studied here, we note that the relative
variations in the radial dependence for the baryon fraction for the
prolate case are much smaller than if the cluster is oblate (see Figures
\ref{fig:z00_densityratio}, \ref{fig:z03_densityratio},
\ref{fig:z06_densityratio}, and \ref{fig:z09_densityratio}).  The shape
of the baryon fraction radial profile provides an interesting
possibility for discriminating between the prolate/oblate hypotheses:
Due to the expected strong correlation between the gas and dark matter
distributions, one expects their density and mass ratios to show
relatively small variability as a function of radius from the cluster
center. Indeed, for low- to intermediate redshift clusters $(z \lesssim
0.5)$, the baryon fraction has a relatively slight radial dependence
$f_b(r) \propto r^{0.2}$ (\cite{ettori99}).

In conclusion, we have demonstrated that: 1) The fundamental assumption
of axial-symmetry is a reasonable assumption when applied to a realistic
subset of the general cluster population. 2) Under the assumption of
axial symmetry, the method works very well for the purpose of recovering
the 3D structure. 3) For a reasonable COI extrapolation scheme, the
deprojection is robust, stable and unique even when most of k-space
information has been lost in the projection. 4) The method is applicable
for deprojecting cluster maps throughout the evolutionary history.

\acknowledgements{We thank A. J. Banday, M. Bartelmann, A. Dekel,
Y. Rephaeli, A. Riess and S. D. M. White for helpful discussions. We
would also like to thank the referee, A. Cooray, for numerous helpful
comments. Support for one of us (GKS) was provided by NASA through
Hubble Fellowship Grant No. HF-01114.01-98A from the Space Telescope
Science Institute, which is operated by the Association of
Universities for Research in Astronomy, Incorporated, under NASA
Contract NAS5-26555. This research has been partially supported by a
US-Israel Binational Science Foundation grant 94-00185 (YH and JS) and
an Israel Science Foundation grant 103/98 (YH).}

\clearpage

\section{Appendix}

The temperature decrement (the Sunyaev-Zel'dovich effect) the
Rayleigh-Jeans regime due to upscattering of microwave background
photons by the diffuse cluster electron gas in is given by
\begin{equation}
\frac{\delta T} { T }(x,y) = -2 y = -2 \frac{\sigma_T}{m_e c^2} 
        \int n_e(\vec{r}) \, k_B \, T_x(\vec{r}) \, dl
\end{equation}
where $\sigma_T$ is the Thompson cross section, $k_B$ the Boltzmann
constant, $m_e$ is the electron mass, and $n_e(\vec{r})$ denotes the 3D
electron gas distribution, with temperature $T_x(\vec{r})$.
 
The X-ray surface brightness in a given passband is given by
\begin{equation}
S_X(x, y) = \frac{1}{4\pi (1+z)^3} \int dl \, n_e^2 (\vec{r})
\Lambda(T_x) 
\end{equation}
where the $\Lambda(T_x)$ is the emissivity which is a function of the
temperature and the energy passband at the cluster redshift. 

These expressions can be rewritten in a general form, factoring out the
normalization and 3D structural forms so that gas density an temperature
are given by $n(x,y,z) = n_0 f_1(\theta,\phi,\omega)$ and $T_x(\vec{r})
= T_{x_0} f_2(\theta,\phi,\omega)$ where $D_A$ is the cluster angular
diameter distance and $l = D_A \omega$. For galaxy clusters, the
dominant emission mechanism is bremsstr\"ahlung in the limit of an
optically thin plasma, so that $\Lambda \propto T_x^{1/2}$ and is a
well-known function (\cite{mewe85}; \cite{mewe86}).  Thus we can write
$\Lambda(T_x) \simeq \Lambda_0(T_{x_0})\, f^{1/2}_2(\theta,\phi,\omega)$
and the SZ decrement can be written
\begin{eqnarray}
\frac{\delta T} { T } & = &   -2 \, D_A \, 
	\frac{k_B T_{X_0}}{m_e c^2} \sigma_T \, n_0 \,
	\int f_1 \, f_2 \,d\omega \nonumber \\
	& \equiv &  \Delta_0 \; Q_{SZ}(f_1,f_2) 
\end{eqnarray}
and
\begin{eqnarray}
S_X & = &  \frac{ n_0^2 \, \Lambda_0 }{ 4 \pi (1+z)^3 } \; D_A \,
   \int f_1^2  \, f_2^{1/2} \, d\omega \\ 
	& \equiv & S_{X_0} \; Q_{X}(f_1, f_2). 
\nonumber
\end{eqnarray}

The combination of the SZ and X-ray observations affords the opportunity
to determine the Hubble constant.  The angular diameter distance to the
cluster is given by
\begin{equation}
D_A =  	\frac{\Lambda_0}{16 \pi (1+z)^3 \sigma_T^2}
	\left( \frac{\Delta_0^2 }{S_{X_0}} \right) 
	\left(  \frac{m_c c^2}{k_B T_{x_0} } \right)^2. \label{eqn:da}
\end{equation}

\subsection{The isothermal $\beta$-model}

Let us consider a concrete, analytical example.  A convenient
parameterization of the gas distribution is given by the standard
$\beta$-model (e.g., \cite{cavaliere76})
\begin{equation}
n_e(r) = n_0 \left( 1 + \frac{r^2}{r^2_c}  \right)^{-3\beta/2}.
\end{equation}

Taking the further simplifying assumption of an isothermal gas, the form
factors are given by
\begin{eqnarray}
Q_{SZ} & = & \sqrt{\pi} \frac{\Gamma(3\beta/2 - 1/2)}{\Gamma(3\beta/2)}
	\theta_c \left( 1 +
	\frac{\theta^2}{\theta_c^2}  \right)^{1/2 - 3 \beta/2}.       
\end{eqnarray}
and 
\begin{eqnarray}
Q_{X} & = & \sqrt{\pi} \frac{\Gamma(3\beta - 1/2)}{\Gamma(3\beta)}
	\theta_c \left( 1 +
	\frac{\theta^2}{\theta_c^2}  \right)^{1/2 - 3 \beta}.
\end{eqnarray}

The required observations are the cluster redshift, a determination of
the gas temperature, central density and a fit for $\beta$ and $\theta_c$.

We can slightly generalize the standard $\beta$-model, to allow for a
more complicated 3D geometry.  Consider as a toy-model a $\beta = 2/3$,
isothermal cluster with a gas density profile given by
\begin{equation}
\rho_g(x,y,z) = \frac{\rho_{g_0} \, r_c^2 }{ r_c^2 + x^2/a^2 + y^2/b^2 +
  z^2/c^2}. 
\end{equation}

Without loss of generality, let us suppose the cluster z-axis forms an angle
$\theta$ with respect to the line of sight.
In this model, the Sunyaev-Zel'dovich decrement is given by
\begin{eqnarray}
\frac{\delta T} { T}& =& -2 \,\pi \,\frac{k_B T_x}{m_e c^2} 
        \sigma_T \,n_0  \,c^\prime \,
        r_c \, \biggl[ 1  +  \frac{x^2}{r_c^2a^2} + 
	\frac{y^2}{r_c^2 (c^2\sin^2\theta +
        b^2\cos^2\theta)}\biggr]^{-1/2} \nonumber  \\
        & = & 2.5 \times 10^{-4} \; c^\prime \;
        \biggl( \frac{ r_c }{ 100 \; {\rm kpc} }\biggr) \;
        \biggl(  \frac{n_0 }{0.01 \; {\rm cm}^{-3} }\biggr) \;
         \biggl(  \frac{T_x }{10 \; {\rm keV} }\biggr) \nonumber \\ 
	& \; & \;\;\;\;\;\;  \times 
         \biggl[ 1  +  \frac{x^2}{r_c^2a^2} + 
\frac{y^2}{r_c^2 (c^2\sin^2\theta + b^2\cos^2\theta)} \biggr]^{-1/2} \nonumber
\end{eqnarray}
where
\begin{eqnarray}
c^{\prime} &=& \biggl(  \sin^2\theta/b^2 + \cos^2\theta/c^2  \biggr)^{-1/2 }.
\end{eqnarray}

The observed surface brightness scales as
\begin{eqnarray}
S_X(x,y) 
        &=&  \frac{ n_0^2 \, \Lambda_0 }{ 4 \pi (1+z)^3 }
	r_c \;
        c^\prime  \biggl[ 1  + \frac{x^2}{r_c^2a^2} + 
        \frac{y^2}{r_c^2(c^2\sin^2\theta + b^2\cos^2\theta) }
        \biggr]^{-3/2} \frac{ {\rm \; erg} }{ {\rm cm}^2 {\rm s} }
\end{eqnarray}

\begin{figure*}[htbn]
\centering \leavevmode
\epsfxsize=0.5\columnwidth \epsffile{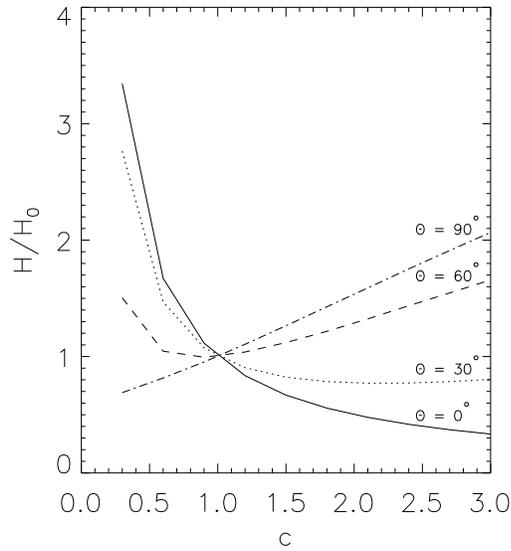}
\caption{The inferred value of the Hubble constant vs. the axis ratio,
c, using the spherical $\beta$ model formalism, when the underlying
surface density is triaxial. The model fit is done using $\chi2$
minimization.}
\label{fig:h0fit}
\end{figure*}

It is immediately clear the error one makes when fitting a spherical
isothermal $\beta$-model when the true underlying gas density follows
this triaxial model. In the special case that $\theta = 0^\circ$, then
from equation (\ref{eqn:da}), we find $\left(\frac{H}{H_0}\right)^{-1}
= c$, where $H$ is the estimated value of the Hubble constant (see
Figure~\ref{fig:h0fit}).

We determine the error in the general case empirically. The value of
$H_0$ extracted from SZ and X-ray observations assuming a spherical
$\beta$ model will depend on the fitted values of the normalizations,
$\beta$ and $r_c$.

\end{document}